\documentclass{article}
\usepackage{arxiv}
\usepackage[utf8]{inputenc} % allow utf-8 input
\usepackage[T1]{fontenc}    % use 8-bit T1 fonts
\usepackage{hyperref}       % hyperlinks
\usepackage{url}            % simple URL typesetting
\usepackage{booktabs}       % professional-quality tables
\usepackage{amsfonts}       % blackboard math symbols
\usepackage{amsmath}
\usepackage{nicefrac}       % compact symbols for 1/2, etc.
\usepackage{microtype}      % microtypography
\usepackage{cleveref}       % smart cross-referencing
\usepackage{graphicx}
\usepackage{natbib}
\usepackage{doi}
\usepackage[title]{appendix}

\title{Unbiased evaluation and calibration of ensemble forecast anomalies}
\author{ \href{https://orcid.org/0000-0002-2958-6637}{\includegraphics[scale=0.06]{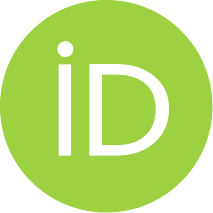}\hspace{1mm}Christopher David Roberts}\\
	ECMWF\\
	Shinfield Park\\
	Reading, United Kingdom\\
	\texttt{chris.roberts@ecmwf.int} \\
	\And
    Martin Leutbecher\\
	ECMWF\\
	Shinfield Park\\
	Reading, United Kingdom\\
}
\renewcommand{\headeright}{PREPRINT}
\renewcommand{\undertitle}{PREPRINT}
\renewcommand{\shorttitle}{Unbiased evaluation and calibration of ensemble forecast anomalies}

\hypersetup{
  colorlinks   = true, %Colours links instead of ugly boxes
  urlcolor     = blue, %Colour for external hyperlinks
  linkcolor    = blue, %Colour of internal links
  citecolor   = red %Colour of citations
}

\begin{document}
\maketitle

%==========================
% A B S T R A C T 
%==========================
\begin{abstract}
Subseasonal, seasonal, and decadal ensemble forecasts are typically presented and verified as anomalies with respect to a lead-time dependent climatological mean to remove the influence of systematic model biases. However, common methods for calculating anomalies result in statistical inconsistencies between forecast and verification anomalies, even for a perfectly reliable ensemble.  It is important to account for these systematic effects when evaluating ensemble forecast systems, particularly when tuning a model to improve the variance and/or reliability of forecast anomalies or when comparing spread-error diagnostics between systems with different reforecast periods. Here, we show that unbiased variances and spread-error ratios can be recovered by deriving estimators that are consistent with the values that would be achieved when calculating anomalies relative to the true, but unknown, climatological mean. An elegant alternative is to construct forecast climatologies separately for each member, which ensures that forecast and verification anomalies are defined relative to reference climatological means with the same sampling uncertainty. This alternative approach has no impact on forecast ensemble means but systematically modifies the total variance and ensemble spread of forecast anomalies in such a way that anomaly-based spread-error ratios are unbiased without any explicit correction for climatology sample size. Furthermore, the improved statistical consistency of forecast and verification anomalies means that probabilistic forecast skill is optimized when the underlying forecast is also perfectly reliable. Alternative methods for anomaly calculation can thus impact probabilistic forecast skill, especially when anomalies are defined relative to climatologies with a small sample size. For example, spread-error ratios derived from 5-year reforecasts can vary by $\pm$12\%, depending on the choice of anomaly calculation method. Finally, we demonstrate the equivalence of anomalies calculated using different methods after applying an unbiased member-by-member statistical calibration. 

\end{abstract}

%==========================
% I N T R O D U C T I O N 
%==========================
\section{Introduction}
\label{section:intro}

Ensemble forecast systems estimate the probability density of the future state of the Earth system and thus provide flow-dependent estimates of forecast uncertainty \citep{leith1974theoretical, hoffman1983lagged, murphy1988impact, molteni1996ecmwf, palmer2005representing, bowler2008mogreps}. The usefulness of such probabilistic forecasts depends on their reliability, which requires that the observed frequency of an event should be equal to the forecast probability when averaged over many events with the same forecast probability \citep{, leutbecher2008ensemble, johnson2009reliability, weisheimer2014reliability}. 

Importantly, and in contrast to medium-range forecasts, long-range forecasts are typically presented and verified as anomalies with respect to a start-date and lead-time dependent climatological mean to remove the influence of systematic model biases. Forecast anomalies are then verified against observed anomalies, which are defined relative to an equivalent observed climatological mean that is sampled to match the available reforecast start dates. This approach to bias correction has a long history and goes back to early studies by \cite{shukla1983physical} and \cite{miyakoda1986one}. 

Real-time forecast anomalies are typically defined relative to the climatological mean of a `reforecast', which comprises a set of forecasts using the same model run for the same (or nearly the same) calendar date in previous years. For example, all models contributing real-time forecast data to the subseasonal-to-seasonal (S2S) prediction project database also provide an associated set of reforecasts \citep{vitart2017subseasonal}. However, there is no consensus on the optimal balance of resources between real-time forecasts and reforecasts, which is reflected in the diversity of reforecast configurations (i.e. ensemble size, reforecast period, frequency of start dates) provided by different S2S modelling groups. Reforecasts are also used to assess model performance, in which case anomalies are defined relative to the climatological mean of the same reforecast. Here, we focus on the evaluation of reforecasts but our results also apply to real-time forecast anomalies (see section \ref{section_method_b}). 

Regardless of the precise configuration of a reforecast, there will always be an element of sampling uncertainty to determine the model climatological mean and associated verification climatological mean \citep[e.g., ][]{leith1973standard, pavia2004uncertainty}. The overarching objective of this study is to consider, given the impacts of this sampling uncertainty, the conditions under which an anomaly-based ensemble forecast system will be perfectly reliable if the underlying forecasts are already perfectly reliable (and thus would not require any bias correction in the first place). This study does not consider the impact of other sources of uncertainty (e.g. observation errors or representativity effects), and thus the underlying `truth' is assumed to be perfectly observed and we use the terms `observation' and `truth' interchangeably. The sampling uncertainty effects discussed in this study have consequences for the evaluation of both perfectly reliable and real-world ensemble forecast systems and are related to the more general question: \emph{what is the optimal method to define forecast and verification anomalies in order to maximise the probabilistic skill of an anomaly-based ensemble forecast system?} 

In a perfectly reliable ensemble forecast, observations and forecast members have the same statistical properties and can be considered drawn from the same underlying probability distribution. This implies that the total variance of forecast members and observations will be equal when evaluated over many events. Following \citet{van2015ensemble}, we refer to this variance equality criterion as \textit{climatological reliability}. There is also a relationship between the expected variance of the ensemble mean error and the expected variance of members about the ensemble mean. \citet{leutbecher2008ensemble} provide an unbiased expression for this relationship, which we reproduce below using their notation.

For a perfectly reliable ensemble of $j=1,\dots,M$ independent cases (e.g. different start dates) and $k=1,\dots,N$ members, the observed state ($x_{\textnormal{o},j}$) and ensemble members ($x_{1,j},\dots,x_{N,j}$) can be considered independent draws from the same probability distribution with mean $\mu_j$ and standard deviation $\sigma_j$. The mean over a sample of ensemble members is represented as $\left<\cdot\right>_{N} \equiv \frac{1}{N}\sum^{N}_{k=1}$, such that the ensemble mean for case $j$ is denoted $\left<x_{.,j}\right>_{N} \equiv \frac{1}{N}\sum^{N}_{k=1}x_{k,j}$. The uncorrected sample variance of the ensemble for case $j$ is given by

\begin{equation}
    \label{eq:ens_variance}
    s_j^2 = \left< \left( x_{.,j} - \left<x_{.,j}\right>_N \right)^2\right>_N,
\end{equation}
and the squared error of the ensemble mean by 

\begin{equation}
    \epsilon_j^2 = \left( x_{\textnormal{o},j} - \left<x_{.,j}\right>_N \right)^2.
\end{equation}

In a perfectly reliable ensemble, the exchangeability of observations ($x_{\textnormal{o},j}$) and forecast members ($x_{k,j}$) implies the following relationships between $\sigma_j$, the expected variance of the ensemble mean error, and the expected variance of members about the ensemble mean

\begin{equation}
    \frac{1}{M} \sum\limits^{M}_{j=1} \left( \frac{N}{N-1} s_j^2 - \sigma_j^2 \right) \rightarrow 0 \textnormal{ as }M \rightarrow \infty,
\end{equation}

\begin{equation}
    \frac{1}{M} \sum\limits^{M}_{j=1} \left( \frac{N}{N+1} \epsilon_j^2 - \sigma_j^2 \right) \rightarrow 0 \textnormal{ as }M \rightarrow \infty, 
\end{equation}
such that 

\begin{equation}
    \frac{1}{M} \sum\limits^{M}_{j=1} \left( \epsilon_j^2 - \frac{N+1}{N-1} s_j^2 \right) \rightarrow 0 \textnormal{ as }M \rightarrow \infty,
\end{equation}

where the factors $\frac{N}{N-1}$, $\frac{N}{N+1}$, and $\frac{N+1}{N-1}$ ensure that these expressions are unbiased with ensemble size and are required because the spread and error refer to the ensemble mean $\left<x_{.,j}\right>$ and not the population mean $\mu_j$. Specifically, $\frac{N}{N-1}$ is Bessel's correction factor that is required for unbiased estimates of the population variance when the population mean is unknown. Similarly, the factor $\frac{N}{N+1}$ accounts for sampling uncertainty in the ensemble mean and is required for estimates of the mean squared error that are unbiased with ensemble size (see appendix A). In other words, in a perfectly reliable ensemble, the average ensemble variance will converge towards the average squared error of the ensemble mean after correction for the finite ensemble size. This relationship can be expressed equivalently in terms of the ratio of the root mean square ensemble spread and root mean square error (RMSE) of the ensemble mean. 

\begin{equation}
    \label{eq:spread_error}
    \frac{\textnormal{Spread}}{\textnormal{RMSE}} = \sqrt{\frac{N+1}{N-1}} \frac{\sqrt{\frac{1}{M}\sum^M_{j=1}s_j^2}}{\sqrt{\frac{1}{M}\sum^M_{j=1}\epsilon_j^2}} \rightarrow 1 \textnormal{ as }M \rightarrow \infty.
\end{equation}      

We refer to this expected relationship between spread and error as \textit{ensemble variance reliability}. The reliability characteristics of operational medium-range forecasts are evaluated routinely at the European Centre for Medium-Range Weather Forecasts \citep[ECWMF; ][]{haiden2023evaluation}, and many other studies have investigated the relationship between spread and error at short- and medium-range lead times \citep[e.g.][]{whitaker1998relationship, yamaguchi2016observation, scherrer2004analysis, hopson2014assessing, rodwell2018flow}. However, there have been comparitatively fewer studies on the relationship between spread and error at subseasonal and seasonal timescales \citep[e.g.][]{barker1991relationship, sun2018subseasonal}. 

The use of ensemble forecast anomalies means that subseasonal and seasonal forecasts are assessed by comparing the average ensemble variance with the average squared error of ensemble mean anomalies. However, the finite sample size of observation and reforecast climatologies means that the process of anomaly calculation (or bias correction more generally) has a systematic impact on estimates of forecast reliability and spread-error ratios. Specifically, anomaly-based estimates of total variance and RMSE are biased for small climatology sample sizes, even for a perfectly reliable ensemble. This is because forecast anomalies are defined relative to a sample climatological mean rather than the true climatological mean. 

Here, we demonstrate methods for calculating forecast anomalies and associated anomaly variance and spread-error ratios that correctly diagnose climatological and ensemble variance reliability of anomalies when the underlying ensemble is perfectly reliable. Section \ref{section:methods} defines four different methods to calculate forecast anomalies and provides unbiased expressions for anomaly-based variance and spread-error ratios. Section \ref{section:results} illustrates the impact of different anomaly calculation methods on (i) spread-error ratios in a perfectly reliable ensemble framework and (ii) probabilistic skill in perfectly reliable and real-world ensemble reforecasts. 

\section{Statistical methods}
\label{section:methods}
\subsection{Ensemble forecast anomaly definitions}
\label{section:anomaly_defs}
This section presents four different approaches (methods A, B, C, and D) to calculating ensemble forecast anomalies, which differ only in the estimation of the reference climatological means. The relevant expressions for unbiased total anomaly variance and unbiased spread-error ratio depend on the chosen method. For clarity, the climatological mean and anomaly definitions in this section assume that forecast start dates are always for the same day of the month but in different years. In this case, there is a single true forecast climatological mean and the number of years in the reforecast is equivalent to the number of forecast start dates ($M$). However, the presented results generalize to data sets that combine anomalies defined relative to different reference climatological means (e.g. for different start dates or locations), provided that all estimated climatologies are calculated using the same number of forecast start dates. 

We consider anomaly-based total variance, ensemble spread, and ensemble mean RMSE estimates to be unbiased for finite values of $M$ if they are statistically consistent with values that would be achieved using the (unknown) forecast and observation population means. We denote forecast and observation anomalies defined relative to these true climatological means using $\alpha_{k,j}$ and $\alpha_{\textnormal{o},j}$, respectively, and define them as follows:

\begin{align}
    \label{eq:alpha}
    \alpha_{k,j} &= x_{k,j} - \mu,  \\
    \label{eq:alpha_obs}
    \alpha_{\textnormal{o},j} &= x_{\textnormal{o},j} - \mu_{\textnormal{o}},
\end{align}
where $\mu$ and $\mu_{\textnormal{o}}$ represent the forecast and observation population means, respectively. 

The four anomaly calculation methods discussed in this paper correspond to different choices regarding the number of years (`\emph{all years}' vs `\emph{other years}') and number of ensemble members (`\emph{all members}' vs `\emph{by member}') that contribute to the reference climatological sample mean for a specific forecast start date. The `\emph{all members}' approaches (methods A and B) define forecast anomalies relative to climatological means that combine information from different start dates and different ensemble members. These approaches are commonly used but result in statistical inconsistencies between forecast and verification anomalies, even for a perfectly reliable ensemble forecast, due to the differing sample sizes of forecast and verification climatologies. In contrast, the `\emph{by member}' approaches (methods C and D) construct forecast climatologies separately for each member, which ensures that forecast and verification anomalies are defined relative to reference climatological means with the same sampling uncertainty. The `\emph{by member}' methods thus provide improved statistical consistency of forecast and verification anomalies such that anomaly-based probabilistic forecast skill is optimized when the underlying forecast is also perfectly reliable.

\subsubsection{Method A: \emph{All-Members-All-Years} climatology} 
%climatological means constructed from all years in the reforecast period}
In this method, forecast anomalies ($a_{k,j}$) and observed anomalies ($a_{\textnormal{o},j}$) are defined relative to a climatological mean defined as the sample mean of all years and members in the reforecast period, including the current year. 

\begin{align}
    a_{k,j} &= x_{k,j} - \frac{1}{M} \sum\limits^{M}_{j=1} \left<x_{.,j}\right>_{N} \\
    a_{\textnormal{o},j} &= x_{\textnormal{o},j} - \frac{1}{M} \sum\limits^{M}_{j=1} x_{\textnormal{o},j}
\end{align}  

The reference climatological mean is the same for all members and therefore the ensemble spread of forecast anomalies is identical to the ensemble spread of raw forecasts. However, the total variance of anomalies and the RMSE of ensemble mean anomalies are biased low compared to values that would be achieved using $\alpha_{k,j}$ and $\alpha_{\textnormal{o},j}$. The impacts of anomaly calculation on variance, RMSE, spread, and spread-error ratios as a function $M$ are illustrated in figure \ref{fig:spread_error_vs_period} for an idealized example based on normally distributed data. These results generalise to data sampled from other distributions, including real ensemble forecast data (see section \ref{section:perfect_ensemble}). 

As discussed in section \ref{section:intro}, \citet{leutbecher2008ensemble} showed that a multiplicative factor of $\sqrt{\frac{N}{N+1}}$ is required for estimates of RMSE that are unbiased with ensemble size, $N$. If anomalies are defined with respect to their true population mean, then the unbiased ensemble mean anomaly RMSE is given by

\begin{equation}
    \textnormal{RMSE}(\alpha_{k,j}, \alpha_{\textnormal{o},j}) = \sqrt { \left( \frac{N}{N+1}  \right) } \sqrt{ \frac{1}{M} \sum\limits^{M}_{j=1} \left( \alpha_{\textnormal{o},j} - \left<\alpha_{.,j}\right>_N\right)^2 }.
\end{equation}
However, in real-world applications, an additional multiplicative factor is required to ensure that estimates are unbiased with the sample size of the reference climatology, $M$. For anomalies calculated using method A, the unbiased estimate of the ensemble mean anomaly RMSE is given by

\begin{equation}
    \textnormal{RMSE}(\alpha_{k,j}, \alpha_{\textnormal{o},j}) = \sqrt { \left( \frac{M}{M-1} \right) \left( \frac{N}{N+1} \right) } \sqrt{ \frac{1}{M} \sum\limits^{M}_{j=1} \left( a_{\textnormal{o},j} - \left<a_{.,j}\right>_N\right)^2 },
\end{equation}
which gives the following unbiased expression for anomaly-based spread-error ratios

\begin{equation}
    \frac{\textnormal{Spread}}{\textnormal{RMSE}} =  \sqrt { \left( \frac{M-1}{M} \right) \left( \frac{N+1}{N-1} \right) } \frac{\sqrt{ \frac{1}{M} \sum^M_{j=1} \left< \left( a_{.,j} - \left<a_{.,j}\right>_N\right)^2\right>_N }} {\sqrt{ \frac{1}{M} \sum^M_{j=1} \left( a_{\textnormal{o},j} - \left<a_{.,j}\right>_N\right)^2 }}.
\end{equation} 
Unbiased estimates of total anomaly variance, where $\textnormal{Var}(\alpha) = \mathbb{E}[\alpha^2]$, are given by

\begin{align}
    \textnormal{Var}(\alpha_{k,j}) &= \mathbb{E}[a_{k,j}^2] + \frac{1}{M-1}\mathbb{E}[\left< a_{.,j}\right>_N^2],\\
    \textnormal{Var}(\alpha_{\textnormal{o},j}) &= \frac{M}{M-1} \mathbb{E}[a_{\textnormal{o},j}^2].
\end{align} 
%<REMOVED_FOR_BREVITY>
% \begin{align}
%     \textnormal{Var}(\alpha_{k,j}) &= \frac{1}{N} \sum^N_{k=1} \left( \frac{1}{M}\sum^M_{j=1}a_{k,j}^2 \right) + \frac{1}{M(M-1)} \sum^M_{j=1} \left< a_{.,j}\right>_N^2,\\
%     \textnormal{Var}(\alpha_{\textnormal{o},j}) &= \frac{1}{M-1} \sum^M_{j=1}a_{k,j}^2.
% \end{align} 
%<REMOVED_FOR_BREVITY>
where the definition for $\textnormal{Var}(\alpha_{\textnormal{o},j})$ is equivalent to the standard expression for unbiased sample variance when the population mean is estimated using the sample mean. In contrast, the appropriate unbiased estimate for $\textnormal{Var}(\alpha_{k,j})$ includes a correction term that depends on the variance of ensemble mean anomalies due to the inclusion of all years and all members in the reference climatological mean.

Importantly, the derivation of these unbiased estimators (see appendices B, C and D) requires that reference climatological means are constructed from $M$ \textit{independent} ensemble mean forecasts and $M$ \textit{independent} observed values. In the case that each forecast is a daily or weekly mean separated by one year, it is justifiable to assume that ensemble mean forecasts are statistically independent. However, this assumption is not valid for climatological means derived from dependent data. For example, real-time forecast anomalies from the operational ECMWF subseasonal ensemble forecasting system are specified relative to a model climate that uses all reforecast dates within a one week window of the current day and month of the real-time forecast. In this case, the use of data from adjacent forecasts means that samples are unlikely to be independent, particularly for shorter lead times, and thus the effective sample size will be less than the total number of forecast start dates used to construct the climatology.  

\subsubsection{Method B: \emph{All-Members-Other-Years} climatology}
\label{section_method_b}
In this method, forecast anomalies ($b_{k,j}$) and observed anomalies ($b_{\textnormal{o},j}$) are calculated relative to climatological means estimated separately for each year in the reforecast period using the data from all other years. This leave-one-year-out approach to the construction of climatological means is commonly used to evaluate reforecasts in a manner that is consistent with the computation of real-time forecast anomalies from a separate reforecast climatology. 

\begin{align}
    b_{k,j} &= x_{k,j} - \frac{1}{M-1} \sum\limits^{M}_{\substack{i=1\\i \neq j}} \left<x_{.,i}\right>_{N}\\
    b_{\textnormal{o},j} &= x_{\textnormal{o},j} - \frac{1}{M-1} \sum\limits^{M}_{\substack{i=1\\i \neq j}} x_{\textnormal{o},i}
\end{align}

As for method A, the reference climatological mean for each start date is the same for all members and thus ensemble spread is unchanged and can be calculated by substituting anomaly values directly into equation \ref{eq:ens_variance}. However, the total variance of anomalies and  ensemble mean anomaly RMSE estimates are biased high compared to values that would be achieved using the true population mean (figure \ref{fig:spread_error_vs_period}). For this definition of forecast anomalies, the unbiased estimate of the ensemble mean anomaly RMSE is given by

\begin{equation}
    \begin{split} 
        \textnormal{RMSE}(\alpha_{k,j}, \alpha_{\textnormal{o},j}) = \sqrt { \left( \frac{M-1}{M} \right) \left( \frac{N}{N+1} \right) } \sqrt{ \frac{1}{M} \sum\limits^{M}_{j=1} ( b_{\textnormal{o},j} - \left<b_{.,j}\right>_N)^2 }
\end{split}
\end{equation}

which gives the following unbiased expression for anomaly-based spread-error ratios

\begin{equation}
    \frac{\textnormal{Spread}}{\textnormal{RMSE}} =  \sqrt { \left( \frac{M}{M-1} \right) \left( \frac{N+1}{N-1} \right) } \frac{\sqrt{ \frac{1}{M} \sum^M_{j=1} \left< ( b_{.,j} - \left<b_{.,j}\right>_N)^2\right>_N }} {\sqrt{ \frac{1}{M} \sum^M_{j=1} \left( b_{\textnormal{o},j} - \left<b_{.,j}\right>_N\right)^2 }}.
\end{equation} 
In the case of real-time forecast anomalies calculated relative to a separate reforecast climatology of sample size $K$, unbiased spread-error ratios are calculated as follows 
\begin{equation}
   \frac{\textnormal{Spread}}{\textnormal{RMSE}}  =  \sqrt { \left( \frac{K+1}{K} \right) \left( \frac{N+1}{N-1} \right) } \frac{\sqrt{ \frac{1}{L} \sum^L_{j=1} \left< ( b_{.,j} - \left<b_{.,j}\right>_N)^2\right>_N }} {\sqrt{ \frac{1}{L} \sum^L_{j=1} ( b_{\textnormal{o},j} - \left<b_{.,j}\right>_N)^2 }}
\end{equation} 
where $L$ is the number of real-time forecasts and $M = K+1$.  As for method A, the climatological reliability of anomalies is not satisfied even when the underlying forecast system is perfectly reliable. Unbiased estimates of the total anomaly variance, where $\textnormal{Var}(\alpha) = \mathbb{E}[\alpha^2]$, are given by

\begin{align}
    \textnormal{Var}(\alpha_{k,j}) &= \mathbb{E}[b_{k,j}^2] - \frac{1}{M}\mathbb{E}[\left< b_{.,j}\right>_N^2], \\
    \textnormal{Var}(\alpha_{\textnormal{o},j}) &= \frac{M-1}{M} \mathbb{E}[b_{\textnormal{o},j}^2].
\end{align}

%<REMOVED_FOR_BREVITY>
% \begin{align}
%     \textnormal{Var}(\alpha_{k,j}) &= \frac{1}{N} \sum^N_{k=1} \left( \frac{1}{M}\sum^M_{j=1}b_{k,j}^2 \right) - \frac{1}{M^2} \sum^M_{j=1} \left< b_{.,j}\right>_N^2,\\
%     \textnormal{Var}(\alpha_{\textnormal{o},j}) &= \frac{M-1}{M^2} \sum^M_{j=1}b_{k,j}^2.
% \end{align}
%<REMOVED_FOR_BREVITY>

Again, these expressions require that reference climatological means are constructed from $M$ \textit{independent} ensemble mean forecasts and $M$ \textit{independent} observed values. Further details are provided in appendices C and D.

\subsubsection{Methods C and D: \emph{By-Member-All-Years} and \emph{By-Member-Other-Years} climatologies}
Methods C and D do not require unbiased estimates of RMSE for unbiased spread-error ratios. Instead, forecast climatological means and anomalies are defined separately for each member to ensure statistical consistency with observed anomalies. Method C is analogous to method A but forecast anomalies ($c_{k,j}$) are defined relative to climatological means that are calculated separately for each member using the sample mean of all years in the reforecast period. Note that ensemble means and observed anomalies are identical to those calculated using method A. 

\begin{equation}
    c_{k,j} = x_{k,j} - \frac{1}{M} \sum\limits^{M}_{j=1} x_{k,j}
\end{equation}

\begin{equation}
    \left<c_{.,j}\right>_{N} \equiv \left<a_{.,j}\right>_{N} 
\end{equation}

\begin{equation}    
    c_{\textnormal{o},j} \equiv a_{\textnormal{o},j} = x_{\textnormal{o},j} - \frac{1}{M} \sum\limits^{M}_{j=1} x_{\textnormal{o},j}
\end{equation}

 Method D is analagous to method B but forecast anomalies ($d_{k,j}$) are defined relative to climatological means that are calculated separately for each member and year in the reforecast period using the data from all other years. Note that ensemble means and observed anomalies are identical to method B. 

\begin{equation}
    d_{k,j} = x_{k,j} - \frac{1}{M-1} \sum\limits^{M}_{\substack{i=1\\i \neq j}} x_{k,i}
\end{equation}

\begin{equation}
    \left<d_{.,j}\right>_{N} \equiv \left<b_{.,j}\right>_{N} 
\end{equation}

\begin{equation}    
    d_{\textnormal{o},j} \equiv b_{\textnormal{o},j} = x_{\textnormal{o},j} - \frac{1}{M-1} \sum\limits^{M}_{\substack{i=1\\i \neq j}} x_{\textnormal{o},i}
\end{equation}

Although ensemble mean anomalies calculated using methods C and D are unchanged relative to methods A and B, the ensemble spread of forecast anomalies is systematically different to the spread of the raw ensemble (figure \ref{fig:spread_error_vs_period}). The expected ensemble variance of the raw forecasts can be expressed in terms of forecast anomalies as  

\begin{equation}
    \begin{split}
    \mathbb{E}[s_j^2] &= \frac{M}{M-1}\mathbb{E}\left[\left< ( c_{.,j} - \left<c_{.,j}\right>_N)^2\right>_N\right] = \frac{M-1}{M}\mathbb{E}\left[\left< ( d_{.,j} - \left<d_{.,j}\right>_N)^2\right>_N\right], \\    
    %<REMOVED_FOR_BREVITY>
    %                  &= \frac{1}{M-1}\sum^M_{j=1} \left< ( c_{.,j} - \left<c_{.,j}\right>_N)^2\right>_N = \frac{M-1}{M^2} \sum^M_{j=1} \left< ( d_{.,j} - \left<d_{.,j}\right>_N)^2\right>_N,   
    %<REMOVED_FOR_BREVITY>
    \end{split}
\end{equation}
which follows logically from the derivations of unbiased MSE presented in appendix C. Therefore, for methods C and D, spread and error change consistently with $M$ such that unbiased spread-error ratios can be calculated using equation \ref{eq:spread_error} without any explicit correction for climatology sample size. Similarly, total anomaly variance estimates from forecast and verification data are also biased but consistent with one another such that climatological reliability is satisfied when the underlying raw ensemble forecast is also perfectly reliable. Unbiased estimates of the total anomaly variance, where $\textnormal{Var}(\alpha) = \mathbb{E}[\alpha^2]$, are given by

\begin{align}
    \textnormal{Var}(\alpha_{k,j}) &=  \frac{M}{M-1} \mathbb{E}[c_{k,j}^2] = \frac{M-1}{M} \mathbb{E}[d_{k,j}^2],\\
    \textnormal{Var}(\alpha_{\textnormal{o},j}) &=  \frac{M}{M-1} \mathbb{E}[c_{\textnormal{o},j}^2] = \frac{M-1}{M} \mathbb{E}[d_{\textnormal{o},j}^2].
\end{align} 

In summary, the anomaly calculation methods introduced above can be divided into two conceptually different categories that differ only in the number of ensemble members that are included in forecast climatological means:

\begin{itemize}
    \item The `\emph{all members}' approaches (methods A and B) are commonly used but result in statistical inconsistencies between forecast and verification anomalies, even for a perfectly reliable ensemble forecast due to the differing sample sizes of forecast and verification climatologies. Anomaly-based estimates of ensemble spread and forecast probabilities are unchanged compared to the raw ensemble forecast. However, for a finite climatology sample size $M$, anomaly-based estimates of RMSE are biased, which results in biased spread-error ratios. Unbiased spread-error ratios can be recovered by defining unbiased estimates of RMSE that would be achieved using $\alpha_{k,j}$ and $\alpha_{\textnormal{o},j}$. Forecast and verification estimates of total anomaly variance are both biased, but not consistent with one another (unless $N=1$). As for RMSE, it is possible to define unbiased estimators for total anomaly variance that are consistent with the values that would be achieved using $\alpha_{k,j}$ and $\alpha_{\textnormal{o},j}$.\\

    \item The `\emph{by member}' approaches (methods C and D) define forecast and verification anomalies in a way that satisfies both climatological and ensemble variance reliability criteria when applied to a perfectly reliable ensemble forecast. These methods define reference climatological means separately for each member such that, for a finite climatology sample size $M$, forecast anomalies have a different ensemble spread to the raw forecasts that is consistent with the biased estimates of RMSE derived from ensemble mean forecast anomalies. The statistically consistent definitions of forecast and verification anomalies imply that spread-error ratios are unbiased, even though spread and RMSE both vary with $M$. Total anomaly variance estimates from forecast and verification data are also biased but consistent with one another such that climatological reliability is satisfied when the underlying raw ensemble forecast is also perfectly reliable.  
\end{itemize}

\subsection{Probabilistic skill evaluation}
\label{section:crps}
To quantify the impact of changes in spread-error ratio on probabilistic forecast skill, we use a homogeneous Gaussian approximation for the continuous ranked probability score (CRPS), which can be expressed in a closed form in terms of the variance of the ensemble mean error, the ensemble variance, and the mean error of the ensemble mean \citep{leutbecher2021understanding}. Specifically, we use equation 12 from \citet{leutbecher2021understanding}, which provides an expression for the expected CRPS that is appropriate for unbiased (i.e. anomaly-based) forecasts

\begin{equation}
    \textnormal{CRPS} = \frac{\epsilon}{\sqrt{\pi}} \left( \sqrt{2 + 2 \sigma_{\ast}^2} - \sigma_{\ast}  \right)
\end{equation}

where $\epsilon$ is equivalent to the anomaly-based RMSE and $\sigma_{\ast}$ is the spread-error ratio as calculated in equation \ref{eq:spread_error}, with both including corrections for ensemble size. 

\subsection{Unbiased ensemble calibration}
\label{section:calibration}

We also evaluate the behaviour of different anomaly calculation methods under the constraint of a simple member-by-member calibration approach that simultaneously enforces both climatological and ensemble variance reliability. Calibrated forecast anomalies ($\hat{z}_{k,j}$) are derived by separately modifying the ensemble mean and perturbations from the ensemble mean as follows

\begin{equation}
    \hat{z}_{k,j} = \kappa \left<z_{.,j}\right>_N + \lambda  \left( z_{k,j} - \left<z_{.,j}\right>_N   \right),
\end{equation}
where $z_{k,j}$ represents anomalies calculated using one of methods A, B, C, or D. This formulation follows \citet{johnson2009reliability} and is the zero-mean forecast anomaly equivalent to the CR+WER calibration for a single parameter described in \citet{van2015ensemble}. The novelty of our implementation is to estimate parameters $\kappa$ and $\lambda$ such that they are unbiased for finite ensemble sizes (figure \ref{fig:calibration_parameters} and appendix  E). This has two important consequences: (1) With adequate training data, calibrated ensemble forecast anomalies derived from all four methods can be perfectly reliable, even for small ensembles. (2) Larger ensemble forecasts (e.g. 50-member real-time forecasts) can be calibrated using estimates of $\kappa$ and $\lambda$ derived from a training set of forecasts with a smaller ensemble size (e.g. 10-member reforecasts). 

\section{Results}
\label{section:results}
\subsection{Spread-error diagnostics in a perfectly reliable ensemble framework}
\label{section:perfect_ensemble}

This section describes the impact of different anomaly calculation methods using a perfectly reliable ensemble constructed from a subset of operational subseasonal reforecasts run during 2021 with the ECMWF Integrated Forecasting System (IFS). The operational subseasonal configuration of the IFS is summarized by \citet{roberts2023euro} and further details of the IFS model are available in the online documentation \citep{ifsdoc}. 

Perfectly reliable ensemble forecasts are constructed from perturbed members 1 to 9 of operational reforecasts for 240 start dates (one per month) for the period 2001-2020. These forecasts are verified against forecast member 10 to provide an idealized framework in which the forecasts are unbiased and are statistically consistent with the specified `truth'. Note that the unperturbed control forecast is excluded as it is not statistically exchangeable with perturbed members. 

\begin{figure}[!htbp]
    \includegraphics[width=15cm]{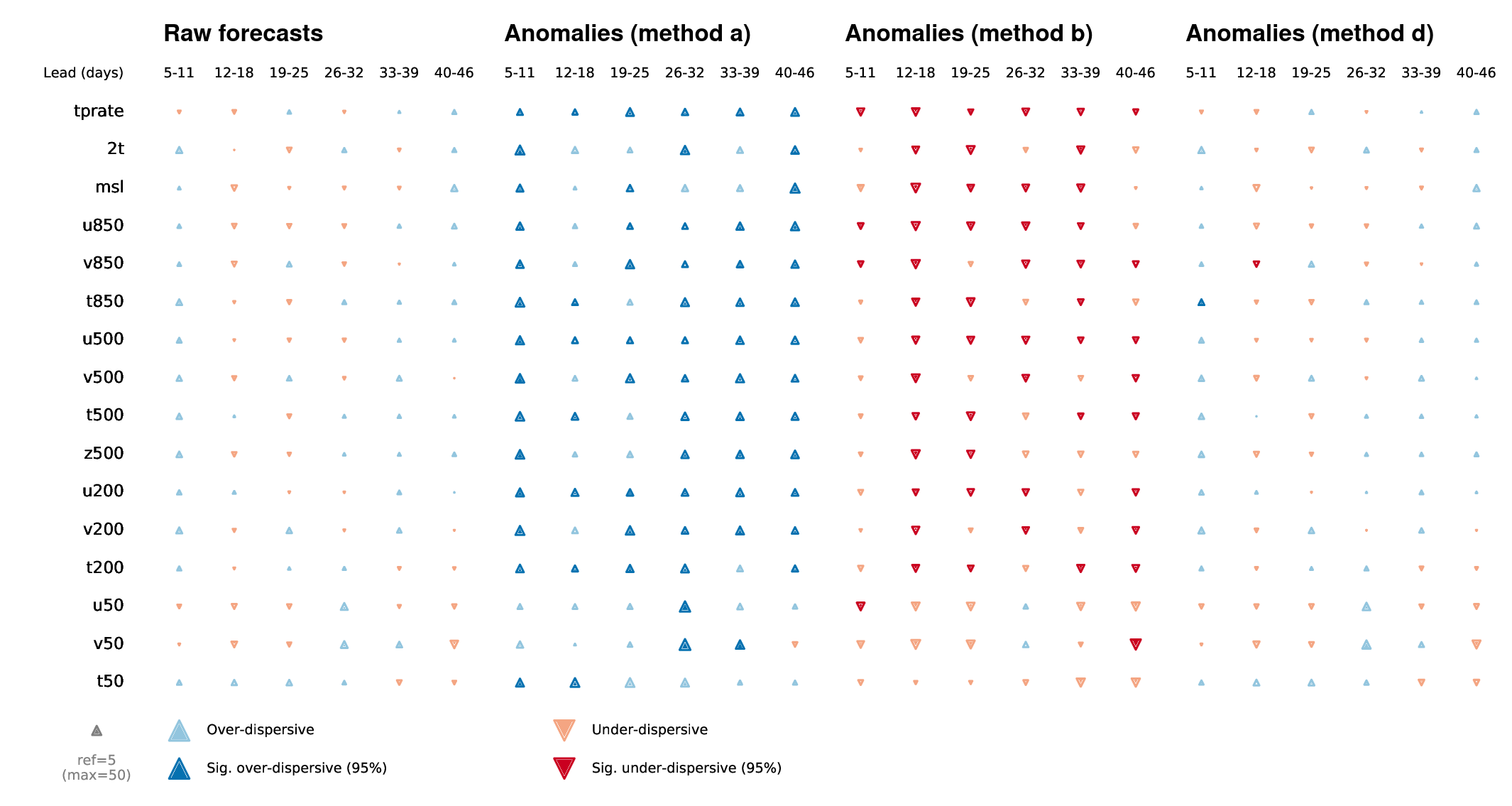}
    \centering
    \caption{Score card summarizing weekly mean spread-error ratios for the northern hemisphere (30$^{\circ}$N-90$^{\circ}$N) in perfectly reliable ensemble reforecasts covering the period 2001-2020 (i.e. $N$=9, $M$=20) as a percentage difference given by $100 \times \left( \left( \frac{\textnormal{Spread}}{\textnormal{RMSE}} \right)_{\textnormal{regional}} - 1 \right)$. All calculations use $\beta_\ast=\sqrt{\frac{N+1}{N-1}}$ (see equation \ref{eq:regional_spread_error}), which produces estimates that are unbiased with ensemble size ($N$) but there is no correction for climatological sample size ($M$). Positive (blue) triangles indicate that spread is larger than RMSE and negative (red) triangles indicate that spread is less than RMSE. Symbol areas are proportional to the magnitude of the percentage difference and significance is determined by bootstrap resampling over start dates. The area of the grey reference triangle corresponds to 5\% difference between spread and error estimates. Spread and RMSE are calculated on a regular 2.5$^{\circ}$ $\times$ 2.5$^{\circ}$ grid for the region north of 30$^{\circ}N$. The variables shown are 2m temperature (2t), total precipitation rate (tprate), mean sea level pressure (msl), temperature (t), zonal/meridional wind (u/v), and geopotential height (z). Numbers in variable names correspond to pressure levels in hPa. }
    \label{fig:spread_error_card_N9_M20_biased}
\end{figure}

We calculate regional spread-skill ratios, using a generalization of equation \ref{eq:spread_error} that includes summation across different locations

\begin{equation}
    \left( \frac{\textnormal{Spread}}{\textnormal{RMSE}} \right)_{\textnormal{regional}} =  \beta_\ast \sqrt{ \frac{ \sum_j \sum_i w_i \left< ( z_{.,j,i} - \left<z_{.,j,i}\right>_N)^2\right>_N } {\sum_j \sum_i w_i ( z_{\textnormal{o},j,i} - \left<z_{.,j,i}\right>_N)^2 } }
    \label{eq:regional_spread_error}
\end{equation}

where $i$ is an index for each model grid box, $w_i$ is a normalized weight to account for variations in grid-box area, $z$ is one of the anomalies $a,b,c,d$,  and $\beta_\ast$ is the appropriate scaling factor, which ensures that estimates are unbiased for small values of $M$ and $N$ for that anomaly. Figure \ref{fig:spread_error_card_N9_M20_biased} summarizes spread-error ratios of weekly means for a nine-member perfectly reliable ensemble covering the period 2001-2020 (i.e. $N$=9, $M$=20).      

There is a good match between spread and RMSE when perfectly reliable ensembles are constructed from raw forecast outputs (figure \ref{fig:spread_error_card_N9_M20_biased}). This result is a consequence of exchangability of ECMWF forecast members such that the perfectly reliable ensemble (perturbed members 1 to 9) is unbiased and statistically interchangeable with the verifying `truth' (perturbed member 10). However, this result is not guaranteed for other forecast systems as members may not be exchangeable if there are systematic differences between members in the forecast model and/or initial perturbations. 

\begin{figure}[!htbp]
    \includegraphics[width=15cm]{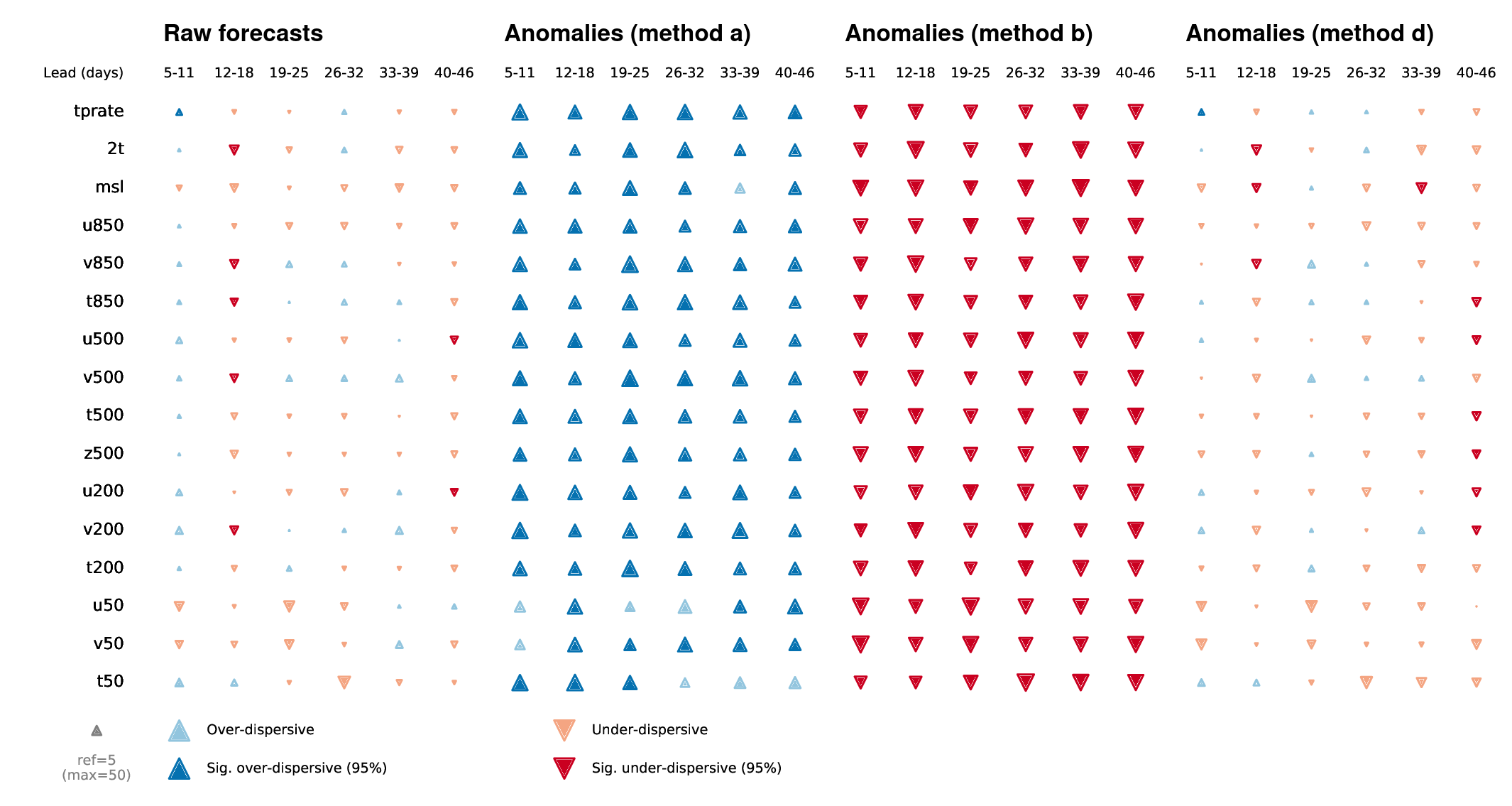}
    \centering
    \caption{Spread-error ratios as figure \ref{fig:spread_error_card_N9_M20_biased}, but for the reforecast period 2016-2020 (i.e. $N$=9, $M$=5).}
    \label{fig:spread_error_card_N9_M5_biased}
\end{figure}

\begin{figure}[!htbp]
    \includegraphics[width=8cm]{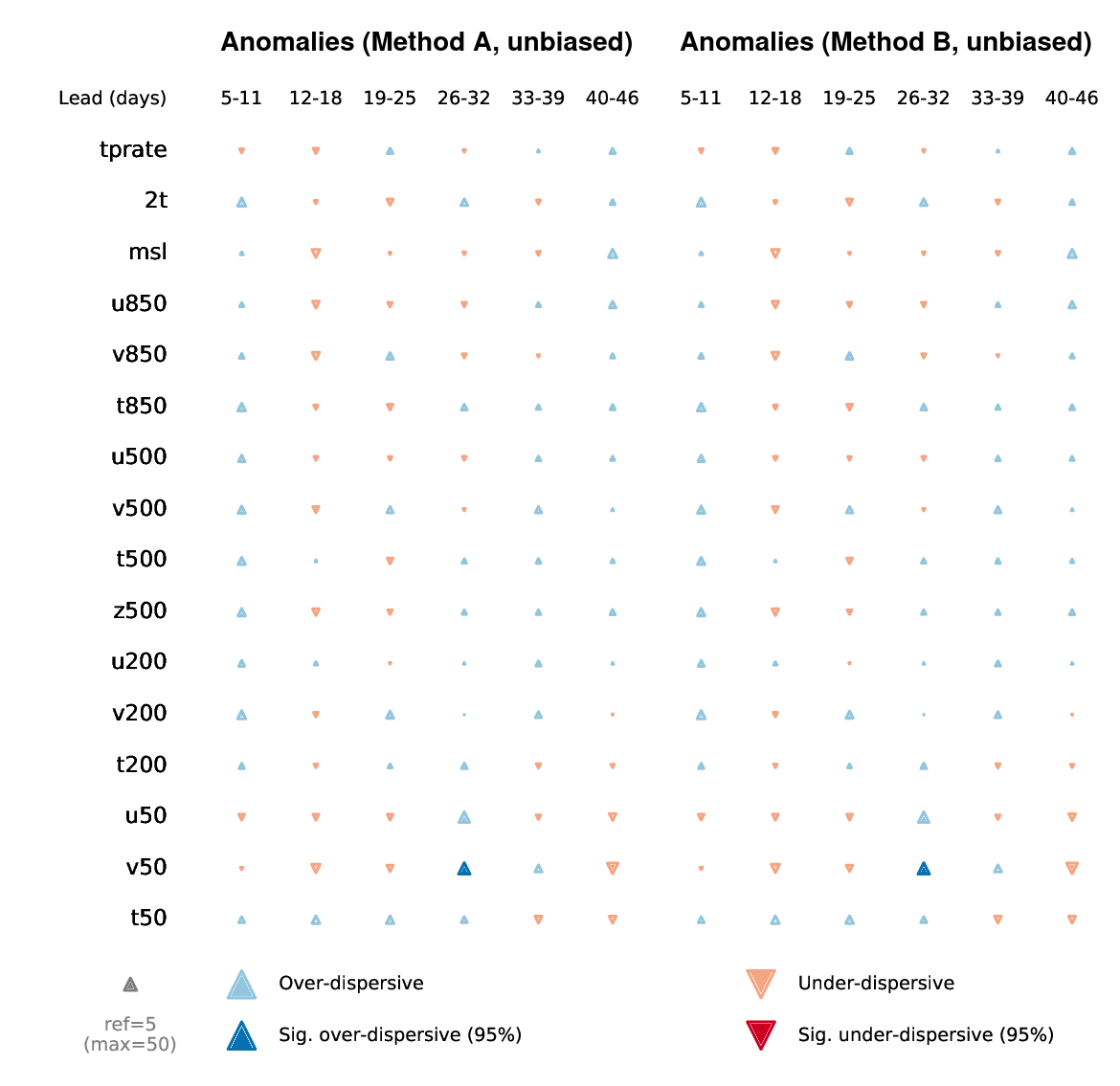}
    \centering
    \caption{Spread-error ratios as figure \ref{fig:spread_error_card_N9_M20_biased}, but calculated using unbiased estimators such that $\beta_{\ast \textnormal{MethodA}}=\sqrt{ \left( \frac{M-1}{M} \right) \left( \frac{N+1}{N-1}\right)}$ and $\beta_{\ast \textnormal{MethodB}}=\sqrt{ \left( \frac{M}{M-1} \right) \left( \frac{N+1}{N-1}\right)}$.}
    \label{fig:spread_error_card_N9_M20_unbiased}
\end{figure}

As expected from the expressions derived in section 2, anomaly-based spread-error ratios calculated using methods A and B are significantly over-dispersive and under-dispersive (figure \ref{fig:spread_error_card_N9_M20_biased}), respectively. This result is more pronounced for a shorter five-year reforecast period (figure \ref{fig:spread_error_card_N9_M5_biased}) and is independent of ensemble size (figure S1). In contrast, anomaly-based spread-error ratios calculated using methods C (not shown) and D are unbiased and consistent with estimates from the raw forecasts (figures \ref{fig:spread_error_card_N9_M20_biased} and \ref{fig:spread_error_card_N9_M5_biased}). For methods A and B, unbiased anomaly-based spread-error ratios can be recovered using the expressions derived in section \ref{section:anomaly_defs} for both 20-year (figure \ref{fig:spread_error_card_N9_M20_unbiased}) and five-year reforecast periods (figure S2).

\subsection{Implications for the probabilistic skill of ensemble forecast anomalies}
\label{section:impacts_on_skill}
In sections \ref{section:anomaly_defs} and \ref{section:perfect_ensemble} it was demonstrated that common approaches to defining forecast anomalies (methods A and B) result in biased variances and spread-error ratios for a perfectly reliable ensemble. These impacts are modest for a typical reforecast period ($\sim$2.5\% difference in spread-error ratio for a 20-year reforecast period) but can be substantial for shorter reforecast periods ($\sim$12\% for a 5-year forecast period). It is therefore important to account for these systematic effects when evaluating new model developments, particularly when tuning the system to improve forecast reliability, including total variance and spread-error ratios. It is especially important to account for such effects when evaluating shorter reforecast configurations or comparing spread-error diagnostics between systems with different reforecast periods. 

For Methods A and B, unbiased variances and spread-error ratios (relative to those achieved using a true population climatological mean) can be recovered with unbiased estimators that account for the sample size of the reference climatology. However, this method is diagnostic only and does not modify total anomaly variance, ensemble spread, or anomaly-based forecast probabilities. It will introduce inconsistencies such that forecast configurations that optimise unbiased anomaly-based spread-error diagnostics are not optimal for other metrics of probabilistic skill, and vice versa.

Furthermore, the unbiased estimators require that reference climatological means are constructed from statistically independent ensemble means (or statistically independent observations). However, real-time forecast anomalies from the operational ECMWF subseasonal ensemble forecasting system are specified relative to a model climate that uses all reforecast dates within a one-week window of the current day and month of the real-time forecast. In this case, the use of data from adjacent forecasts means that samples are unlikely to be independent.
    
An elegant alternative is to construct forecast anomalies separately for each member (methods C and D), which ensures that forecast and verification anomalies are defined relative to reference climatological means with the same sampling uncertainty. This approach has no impact on forecast ensemble means but systematically modifies ensemble variance in such a way that anomaly-based spread-error ratios are unbiased without any explicit correction for climatology sample size. Futhermore, in a perfectly reliable ensemble scenario, methods C and D provide more reliable anomaly forecasts due to the improved statistical consistency of forecast and verification anomalies. 

\begin{figure}[!htbp]
    \includegraphics[width=8cm]{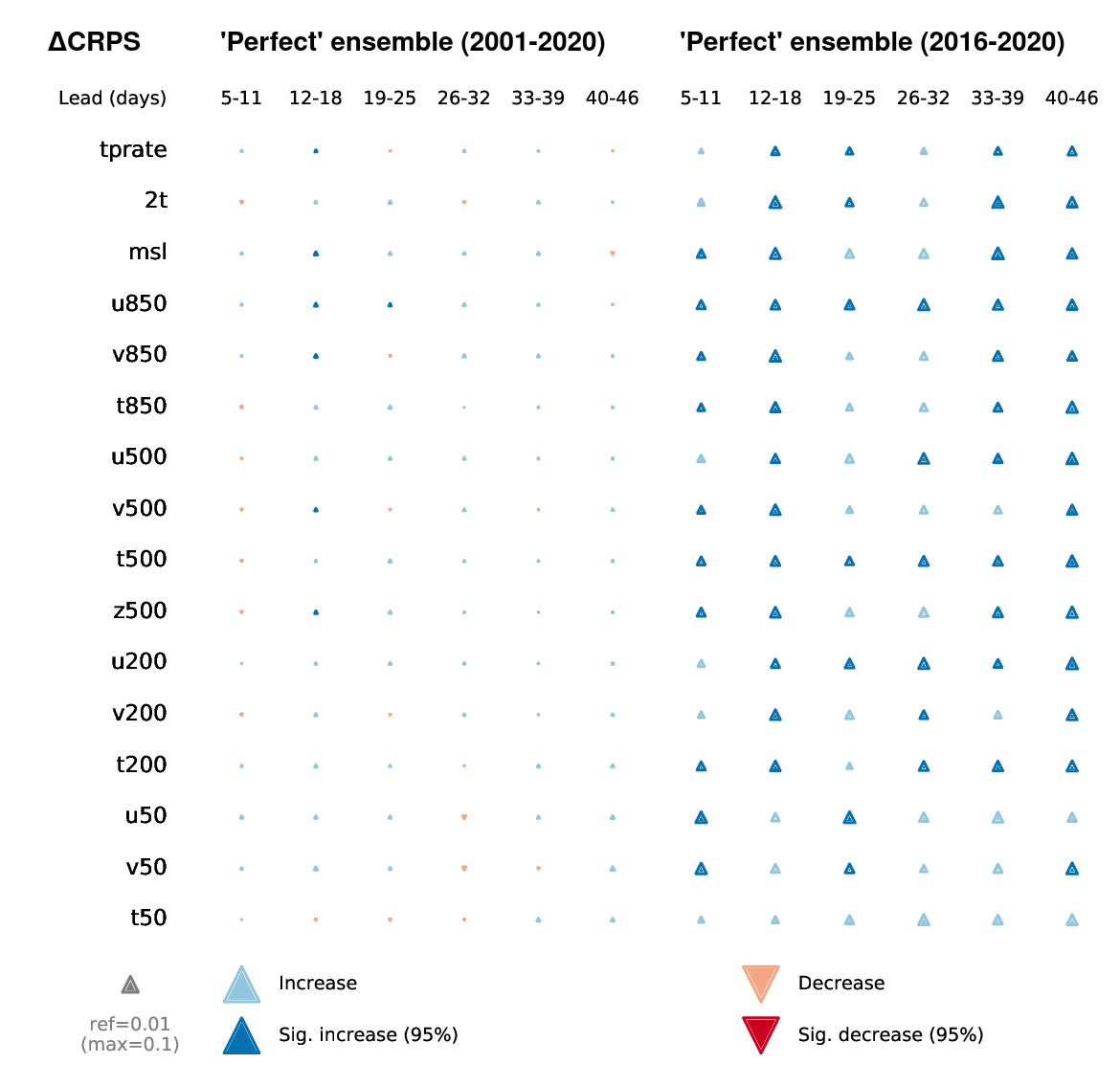}
    \centering
    \caption{Score card summarizing $\Delta CRPS = 1 - \frac{CRPS_{MethodD}}{CRPS_{MethodB}}$ for weekly mean anomalies calculated using methods D and B. Scores are averaged over the the northern hemisphere using data from the nine-member perfectly reliable ensemble reforecasts described in section \ref{section:perfect_ensemble} and cover the periods 2001-2020 or 2016-2020, as indicated in the titles. Positive (blue) triangles indicate that CRPS is improved when anomalies are calculated using method D and the symbol areas are proportional to the magnitude of the difference. The area of the grey reference triangle corresponds to value of 0.01. Significance is determined by bootstrap resampling over start dates. Variable names are as defined in figure 2.}
    \label{fig:crps_perfect_ens}
\end{figure}

\begin{figure}[!htbp]
    \includegraphics[width=8cm]{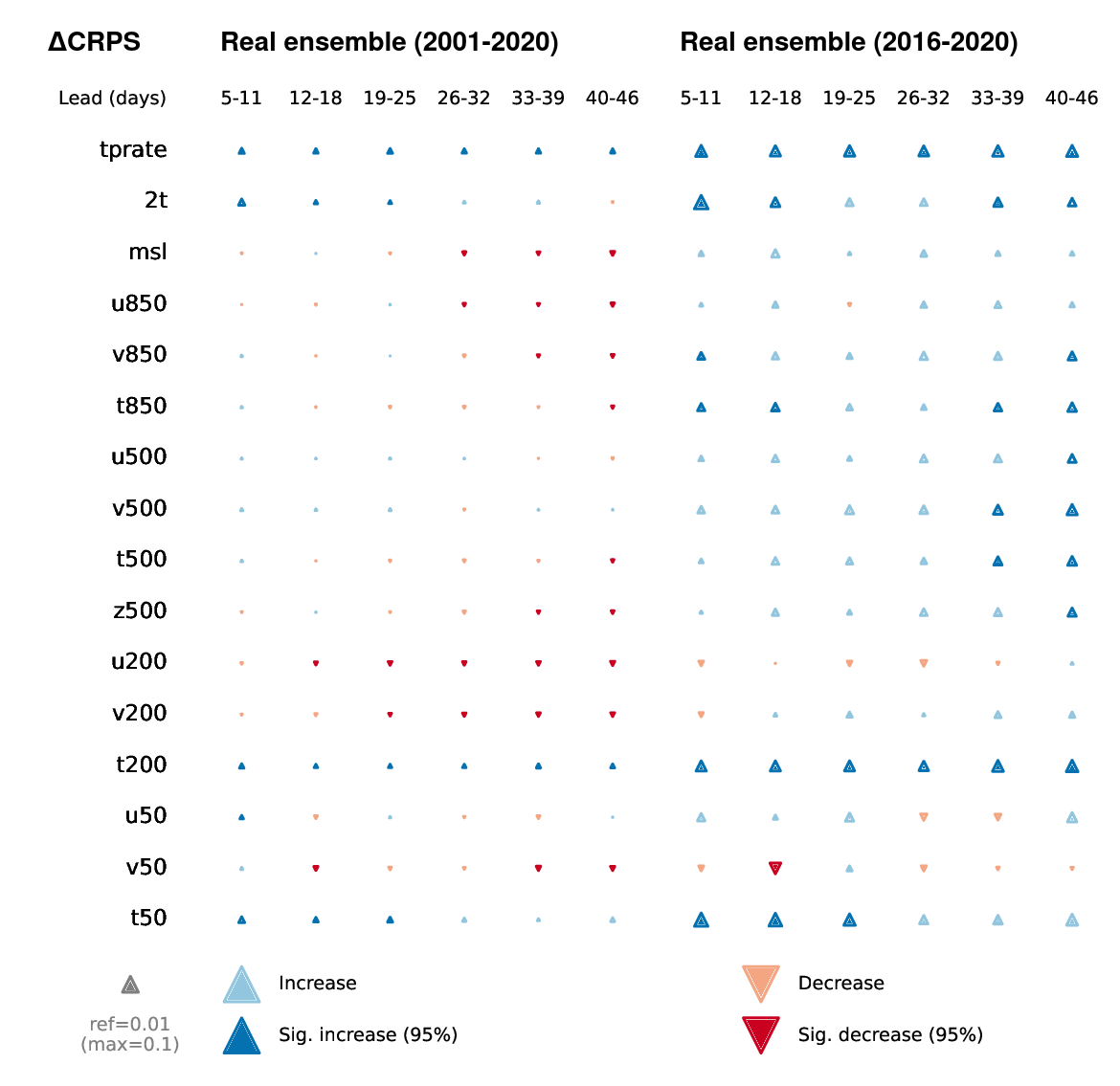}
    \centering
    \caption{As figure \ref{fig:crps_perfect_ens}, but verifying the same ECMWF ensemble reforecasts against the ERA5 reanalysis \citep{hersbach2020era5}.}
    \label{fig:crps_real_ens}
\end{figure}

The impacts on probablistic skill in a perfectly reliable ensemble are illustrated in figure \ref{fig:crps_perfect_ens}, which summarizes differences in CRPS for anomalies calculated using methods B and D. For a 20-year reforecast period, the differences in CRPS are extremely small when the underlying ensemble is perfectly reliable. The impacts are clearer in a 5-year reforecast, where the unbiased spread-error ratio using method D translates to small ($<$ 0.5\%) but consistent improvements to CRPS (figure \ref{fig:crps_perfect_ens}). These results do not translate directly to real reforecasts verified against reanalysis data (figure \ref{fig:crps_real_ens}). Although some variables (e.g. 200 hPa temperature) show evidence for improvements to CRPS using method D compared to method B, other variables are degraded. This is because the underlying forecasts are not perfectly reliable. 

\begin{figure}[!htbp]
    \includegraphics[width=12cm]{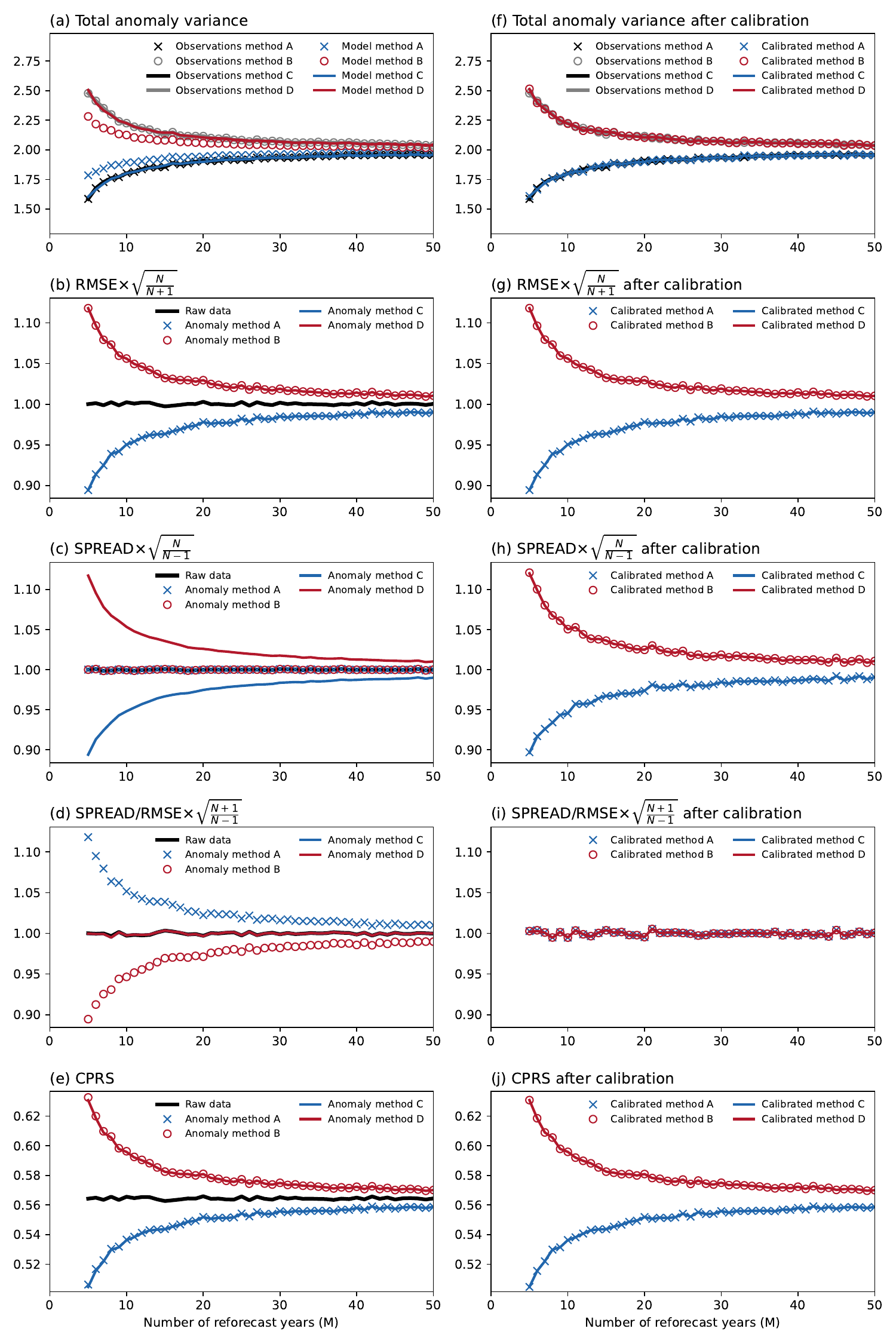}
    \centering
    \caption{(a) Total anomaly variance, (b) ensemble mean RMSE, (c) mean ensemble spread, (d) spread-error ratio, and (e) CRPS, calculated from an idealized perfectly reliable ensemble data set for raw forecasts and different methods of anomaly calculation. Synthetic model and observation data are generated by the same process such that $x_{k,j} = s_{j} + n_{k,j}$, where $s_{j} \sim \mathcal{N}(10,\,1^{2})$ is a predictable component common to all members and observations and $n_{k,j} \sim \mathcal{N}(0,\,1^{2})$ is an unpredictable noise component. The presented values are calculated for a range of reforecast periods ($M=5,...,50$) using N=10 members and averaged over 10000 independent locations. (f-j) As a-e, but calibrated using the unbiased member-by-member calibration approach described in appendix E trained on an independent idealized dataset of the same dimensions generated using the same process.}
    \label{fig:spread_error_vs_period}
\end{figure}

\begin{figure}[!htbp]
    \includegraphics[width=12cm]{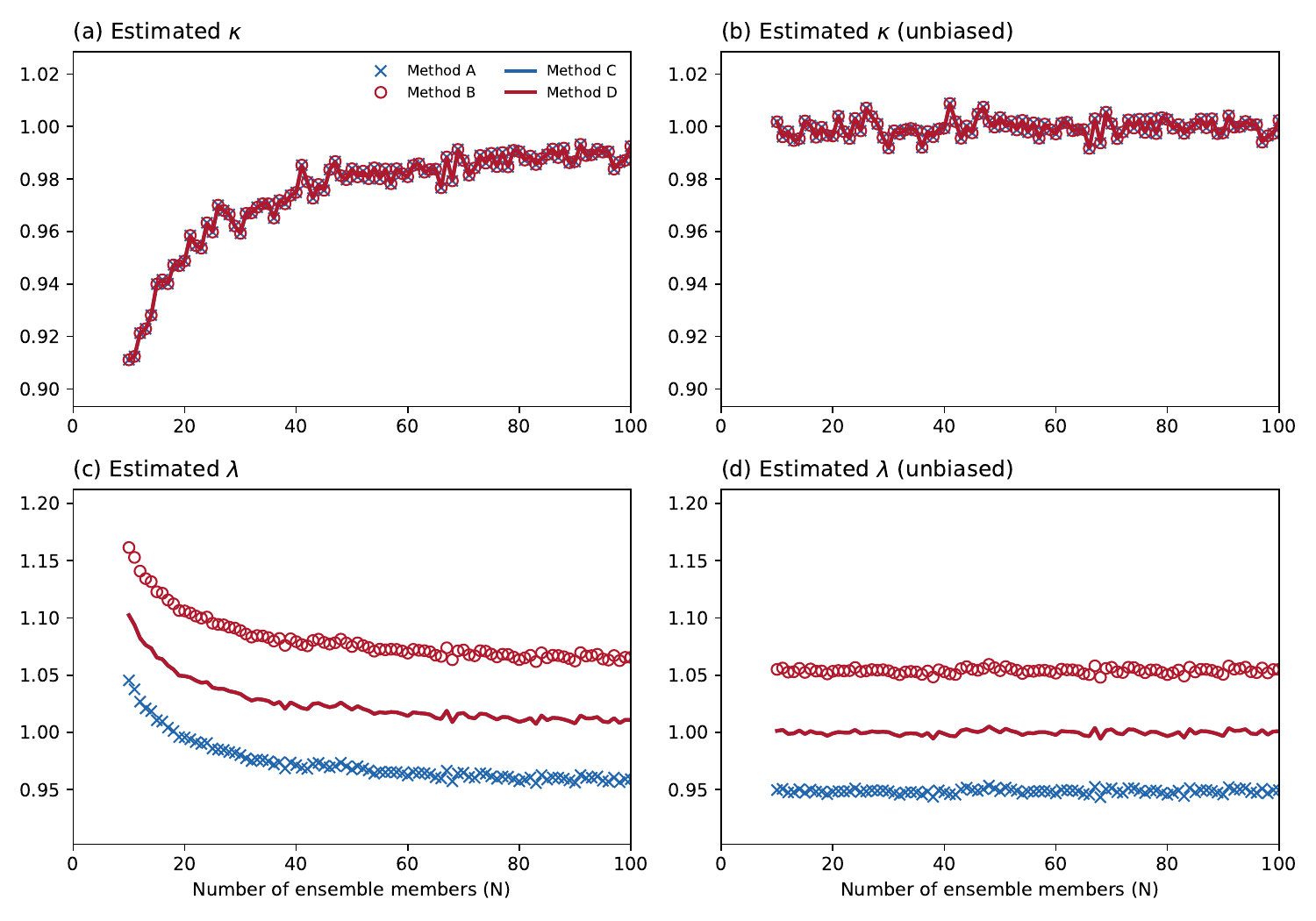}
    \centering
    \caption{Member-by-member calibration parameters for different methods of anomaly calculation applied to an idealized perfectly reliable ensemble data set. The idealized data are generated by the same process illustrated in  figure \ref{fig:spread_error_vs_period} but with a fixed reforecast period ($M=10$) and various ensemble sizes ($N=10,..,100$). (a, c) Parameters $\kappa$ and $\lambda$ are estimated following \citet{johnson2009reliability} and are biased for small ensembles due the increase of $\textnormal{Corr}(\left<z_{.,j}\right>_N, z_{\textnormal{o},j})$ with increasing ensemble size. (b,d) Unbiased estimates of $\kappa$ and $\lambda$ calculated following appendix E, which are insensitive to ensemble size and estimate the expected values ($\kappa \approx 1$ and $\lambda \approx 1$) when forecast anomalies are statistically exchangeable with observed anomalies (Methods C and D).}
    \label{fig:calibration_parameters}
\end{figure}

To further illustrate the impact of anomaly calculation methods on probabilistic skill, we calculate CRPS for idealized data with different spread-error properties. When the underlying raw forecasts are perfectly reliable (figure \ref{fig:spread_error_vs_period}), differences in anomaly-based CRPS are dominated by differences in RMSE. Specifically, methods A and C have identical ensemble means and RMSE is biased low, which results in lower values of CRPS. Similarly, methods B and D have identical ensemble means and RMSE is biased high, which results in higher values of CRPS. The systematic differences in anomaly-based RMSE and CRPS are a consequence of the inclusion (methods A and C) or exclusion (methods B and D) of raw  data for year $j$ in the reference climatological mean used to calculate anomalies for year $j$. Importantly, forecast skill quantified using anomaly methods A/C should not be compared directly with estimates quantified using anomaly methods B/D. 

Our results from idealized synthetic data are consistent with the results based on a perfectly reliable ensemble constructed from ECMWF forecasts. When the underlying raw forecasts are perfectly reliable, the impact of biased spread-error ratios for methods A and B on CRPS are small ($<$ 0.5\%) compared the impact of differences in RMSE between methods A/C and D/B (figure \ref{fig:spread_error_vs_period}). The differences between methods A and C in CRPS are sufficiently small that it is not possible to distinguish a difference between the plotted symbols in figure \ref{fig:spread_error_vs_period}e. The same is true for methods B and D. 

\begin{figure}[!htbp]
    \includegraphics[width=12cm]{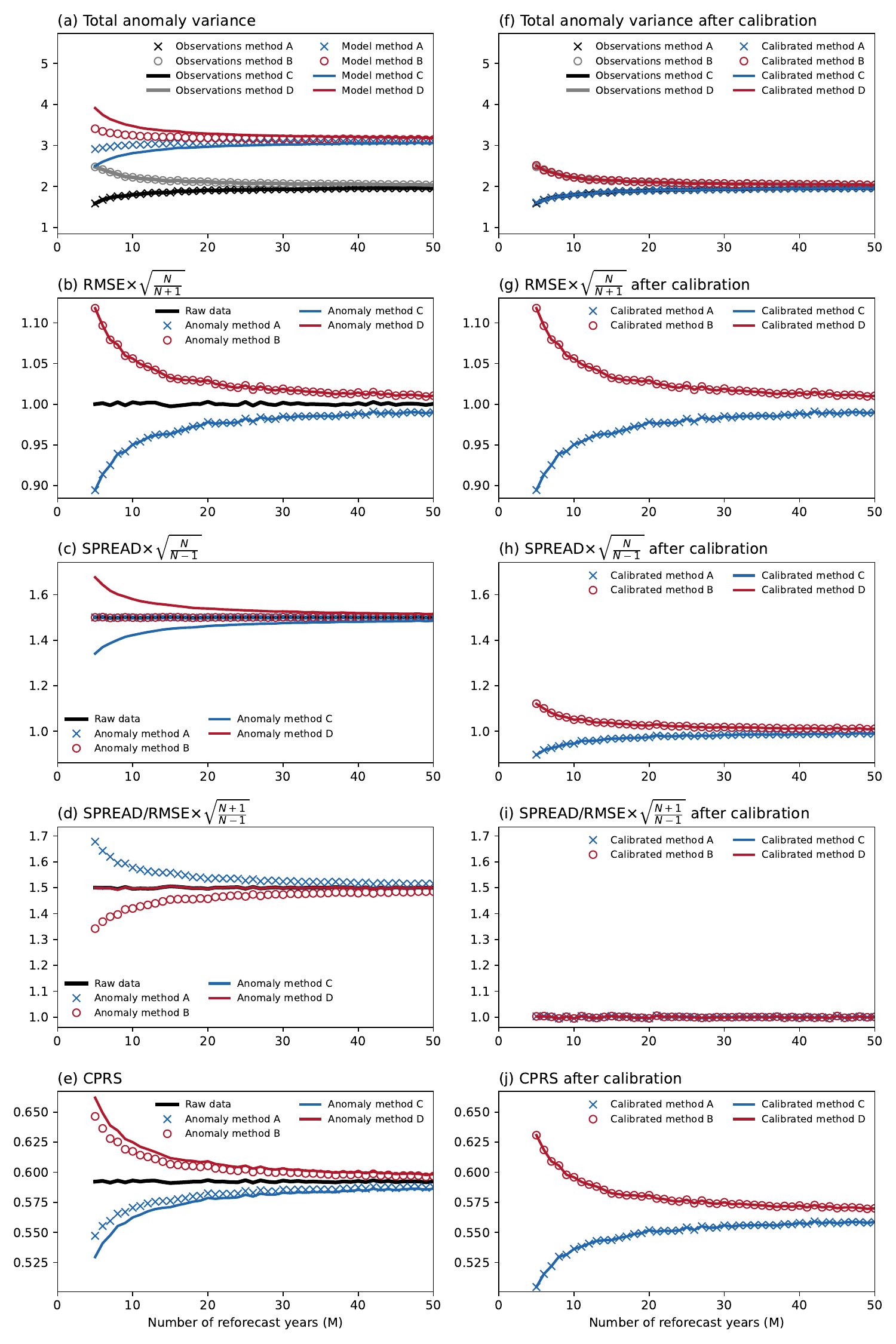}
    \centering
    \caption{As figure \ref{fig:spread_error_vs_period}, but for an overdispersive forecast in which ensemble spread is inflated by a factor of 1.5.}
    \label{fig:spread_error_vs_period_overdispersive}
\end{figure}

The effects of spread differences are clearer when the underlying forecasts are not reliable. For an overdispersive forecast system (i.e. spread $>$ RMSE; figure \ref{fig:spread_error_vs_period_overdispersive}), the lower total variance and reduced ensemble spread for anomalies calculated using method B relative to method D results in a reduction of CRPS. Similarly, the higher variance and ensemble spread for anomalies calculated using method A relative to method C results in an increase of CRPS. These impacts are reversed when the underlying forecast system is underdispersive  (i.e. spread $<$ RMSE; not shown). Importantly, the impact on CRPS of a chosen anomaly calculation method depends on the reliability of the underlying forecasts. However, only for methods methods C and D will anomaly-based CRPS be optimised when the underlying forecast is also perfectly reliable. 

Statistically consistent definitions of forecast and verification anomalies (e.g. by using a single member to calculate reference climatological means) thus have the potential to have modest but positive impacts for operational real-time anomaly forecasts. However, any operational implementation of anomaly calculation methods C and D would need to account for the different ensemble size of real-time forecasts and reforecasts. For example, forecast climatological means could be constructed by randomly drawing a single member from each year of the reforecast period. However, this would add significant complexity to the generation of anomaly-based forecast products.

\subsection{Impacts of calibration}
\label{section:impacts_of_calibration}
Finally, we use idealized data to illustrate the equivalence of anomaly calculation method pairs (A, C) and (B, D) following calibration with an unbiased member-by-member approach (see section \ref{section:calibration}). This calibration is a generalisation of the approach described by \citet{johnson2009reliability} and results in calibrated forecasts with unbiased spread-error ratios, even for small ensemble sizes. This is achieved by constraining calibrated forecasts using the same climatological reliability criterion but enforcing an unbiased spread-error ratio for finite ensemble sizes instead of a correlation-based constraint.

When the underlying ensemble is perfectly reliable, calibration has no impact (i.e. $\kappa \approx 1$ and $\lambda \approx 1$) on anomalies calculated using methods C and D as they are already statistically exchangeable with observed anomalies calculated using the equivalent methods (figure \ref{fig:spread_error_vs_period}). In constrast, calibration modifies the total variance and ensemble spread of methods A and B such that they are equivalent to those calculated using methods C and D, respectively. If the underlying ensemble is not reliable, calibration enforces climatological and ensemble variance reliability constraints for all anomaly calculation methodologies (figure \ref{fig:spread_error_vs_period_overdispersive}). Again, methods A and B become statistically exchangeable with methods C and D, respectively.

\section{Discussion and conclusions}
\label{section:conclusions}

In this study it has been shown that common approaches to calculating ensemble forecast anomalies (methods A and B defined in section \ref{section:anomaly_defs}) result in statistical inconsistencies between forecast and verification anomalies, even for a perfectly reliable ensemble, due to the differing sample sizes of forecast and verification climatologies. For a finite climatology sample size of $M$, this lack of exchangeability between forecast and verification anomalies results in biased estimates of total anomaly variance, anomaly-based ensemble mean RMSE, and anomaly-based spread-error ratios. We have illustrated these concepts using idealized data and operational output from the ECMWF subseasonal forecasting system. However, the results are general and also relevant for other anomaly-based ensemble forecasting systems, such as those used for seasonal and decadal forecasting \citep[e.g., ][]{meehl2021initialized}. 

These systematic effects should be accounted for when defining and evaluating forecast anomalies, particularly when tuning a forecast system to improve the variance and/or reliability of forecast anomalies or when comparing spread-error diagnostics between systems with different reforecast periods. Unbiased approaches are particularly important when it is impossible or impractical to run very long reforecasts. Example scenarios where unbiased approaches might be particularly important include (1) the evaluation of prototype model configurations that are computationally very expensive, (2) efficient testing of many candidate model configurations with a cheap reforecast configuration, and (3) assessment of data-driven forecasting approaches that must be evaluated using reforecast periods that are not within the model training period.

For anomaly calculation methods A and B, unbiased variance and spread-error estimates can be recovered using estimators that are consistent with the values that would be achieved using the true climatological mean. An elegant alternative is to construct forecast anomalies separately for each member (methods C and D in section 2), which ensures that forecast and verification anomalies are defined relative to reference climatologies with the same sampling uncertainty. This alternative approach has no impact on forecast ensemble means but systematically modifies the ensemble variance for forecast anomalies in such a way that anomaly-based spread-error ratios are unbiased without any explicit correction for climatology sample size. The improved statistical consistency of forecast and verification anomalies can have modest but positive impacts on the probabilistic skill of real-time anomaly-based forecast products, especially those with short reforecast periods. Crucially, only methods C and D will provide optimal anomaly-based probabilistic skill when the underlying forecast is also perfectly reliable. 

Finally, we demonstrate that ensemble forecast anomalies calculated using all four methods can be perfectly reliable following calibration with an unbiased member-by-member approach, provided there is sufficient training data to estimate the required parameters. This calibration enforces unbiased climatological and ensemble variance reliability constraints such that methods A and B become statistically exchangeable with methods C and D, respectively. 

\section*{acknowledgements}
We thank Frederic Vitart, Peter Dueben, Llorenç Lledo, and two anonymous reviewers for their constructive feedback on earlier versions of this manuscript. The operational ECMWF reforecast data used in this study are available from the S2S database \citep{vitart2017subseasonal} from \url{https://apps.ecmwf.int/datasets/data/s2s}.

\section*{conflict of interest}
The authors declare no conflict of interest.

%==========================
%   A P P E N D I C E S 
%==========================
\appendixpage
\begin{appendices}

\section{The expected mean squared error (MSE) of a perfectly reliable ensemble forecast}
\citet{leutbecher2008ensemble} showed that a multiplicative factor of $\frac{N}{N+1}$ is required for estimates of the mean squared error (MSE) that are unbiased with ensemble size, $N$. Consider the expected MSE of the ensemble mean from a perfectly reliable ensemble, in which ensemble members $x_{1, j},\dots,x_{N,j}$ and the true state $x_{\textnormal{o},j}$ are independent draws from the same distribution with mean $\mu_j$ and variance $\sigma^2_j$ such that

\begin{equation}
    \begin{split}
        \textnormal{MSE} &= \mathbb{E}[(\left< x_{.,j} \right>_N - x_{\textnormal{o},j})^2], \\
                         &= \mathbb{E}[\left< x_{.,j} \right>_N^2] + \mathbb{E}[x_{\textnormal{o},j}^2] - 2\mathbb{E}[\left< x_{.,j} \right>_N x_{\textnormal{o},j}], \\
                         %<REMOVED_FOR_BREVITY>
                         %&= \frac{1}{N}\sigma^2 + \mu^2 + \sigma^2 + \mu^2 - 2\mu^2, \\
                         %<REMOVED_FOR_BREVITY>
                        &= \frac{N+1}{N}\sigma_j^2, 
    \end{split}
\end{equation}
where the independence of forecast members and observations means that $\mathbb{E}[\left< x_{.,j} \right>_N x_{\textnormal{o},j}] = \mathbb{E}[\left< x_{.,j} \right>_N]\mathbb{E}[x_{\textnormal{o},j}] = \mu_j^2$. From this result it is evident that a multiplicative factor of $\frac{N}{N+1}$ is required to provide an unbiased estimate of the MSE that would be achieved for $N\rightarrow \infty$.

\section{Relationship between anomalies calculated using different methods}
Ensemble mean anomalies calculated using methods A and B are related as follows:

\begin{equation}
    \begin{split}
        \left<a_{.,j}\right>_{N} -  \left<b_{.,j}\right>_{N} &=  \frac{1}{M-1} \sum\limits^{M}_{\substack{i=1\\i \neq j}} \left<x_{.,i}\right>_{N} - \frac{1}{M} \sum\limits^{M}_{i=1} \left<x_{.,i}\right>_{N} \\
        %<REMOVED_FOR_BREVITY>
        % &=  \frac{1}{M-1} \left[  \left( \sum\limits^{M}_{i=1} \left<x_{.,i}\right>_{N} \right) - \left<x_{.,j}\right>_{N}  \right] - \frac{1}{M} \sum\limits^{M}_{i=1} \left<x_{.,i}\right>_{N} \\
        % &=  \frac{1}{M(M-1)}\sum\limits^{M}_{i=1} \left<x_{.,i}\right>_{N}  - \frac{1}{M-1}   \left<x_{.,j}\right>_{N} \\
        %</REMOVED_FOR_BREVITY>
        &=  \frac{1}{M-1} \left[ \left( \frac{1}{M} \sum\limits^{M}_{i=1} \left<x_{.,i}\right>_{N} \right) - \left<x_{.,j}\right>_{N} \right]. 
    \end{split}
\end{equation}
This implies that anomalies using method B can be expressed as
%<REMOVED_FOR_BREVITY>
% \begin{equation}
%     \begin{split}
%         \left<b_{.,j}\right>_{N} &= \left<a_{.,j}\right>_{N} + \frac{1}{M-1} \left(\left<x_{.,j}\right>_{N}  - \frac{1}{M} \sum\limits^{M}_{i=1} \left<x_{.,i}\right>_{N} \right) \\
%         &= \left<a_{.,j}\right>_{N} + \frac{1}{M-1} \left<a_{.,j}\right>_{N} \\
%         &= \frac{M}{M-1} \left<a_{.,j}\right>_{N}.
%     \end{split}
% \end{equation}
%<REMOVED_FOR_BREVITY>
\begin{equation}
    \left<b_{.,j}\right>_{N} = \frac{M}{M-1} \left<a_{.,j}\right>_{N}.
\end{equation}
Similarly, ensemble member anomalies using method B can be expressed as
%<REMOVED_FOR_BREVITY>
% \begin{equation}
%     \begin{split}
%         b_{k,j} &= a_{k,j} + \frac{1}{M-1} \left<a_{.,j}\right>_{N} \\
%                 &= a_{k,j} + \frac{1}{M} \left<b_{.,j}\right>_{N}.
%     \end{split}
% \end{equation}
%</REMOVED_FOR_BREVITY>
\begin{equation}
        b_{k,j} = a_{k,j} + \frac{1}{M} \left<b_{.,j}\right>_{N}.
\end{equation}

\section{Unbiased anomaly MSE estimates}
Here, we derive anomaly-based estimates of MSE that are unbiased with the sample size of the reference climatology, $M$, provided reference climatological means are constructed from $M$ \textit{independent} ensemble mean forecasts and $M$ \textit{independent} observed values. We start from equations \ref{eq:alpha} and \ref{eq:alpha_obs}, which define unbiased forecast anomalies ($\alpha_{k,j}$) and unbiased observed anomalies ($\alpha_{\textnormal{o},j}$) relative to their true population means. The unbiased anomaly mean squared error (MSE) can then be expressed as the expectation of the squared difference between ensemble mean forecast anomalies and observed anomalies.

\begin{align}
    \alpha_{k,j} &= x_{k,j} - \mu\\
    \alpha_{T,j} &= x_{T,j} - \mu _T
\end{align}
The unbiased anomaly mean squared error (MSE) can then be expressed as the expectation of the squared difference between ensemble mean forecast anomalies and observed anomalies. 
\begin{equation}
    \textnormal{MSE}(\alpha_{k,j},\alpha_{\textnormal{o},j}) = \left( \frac{N}{N+1} \right) \mathbb{E}[(\left< \alpha_{.,j} \right>_N - \alpha_{\textnormal{o},j})^2] ,
\end{equation}
which can then be expanded as follows

%</REMOVED_FOR_BREVITY>
% \begin{equation}
%     \begin{split}
%         \textnormal{MSE}(\alpha_{k,j},\alpha_{\textnormal{o},j})&= \left( \frac{N}{N+1} \right) \left(  \mathbb{E}[\left< \alpha_{.,j} \right>_N^2] + \mathbb{E}[\alpha_{\textnormal{o},j}^2] - 2\mathbb{E}[\left< \alpha_{.,j} \right>_N \cdot \alpha_{\textnormal{o},j}] \right)\\
%                         &= \left( \frac{N}{N+1} \right) \left(  \mathbb{E}[\left< \alpha_{.,j} \right>_N^2] + \mathbb{E}[\alpha_{\textnormal{o},j}^2] - 2 \textnormal{Cov}(\left< \alpha_{.,j} \right>_N, \alpha_{\textnormal{o},j}) - 2 \mathbb{E}[\left< \alpha_{.,j} \right>_N] \mathbb{E}[\alpha_{\textnormal{o},j}] \right)\\
%                         &= \left( \frac{N}{N+1} \right) \left(  \mathbb{E}[\left< \alpha_{.,j} \right>_N^2] + \mathbb{E}[\alpha_{\textnormal{o},j}^2] - 2\sqrt{\mathbb{E}[\left< \alpha_{.,j} \right>_N^2]\mathbb{E}[\alpha_{\textnormal{o},j}^2]} \textnormal{Corr}(\left< \alpha_{.,j} \right>_N, \alpha_{\textnormal{o},j}) \right)
%     \end{split}
% \end{equation}
%</REMOVED_FOR_BREVITY>

\begin{equation}
    \textnormal{MSE}(\alpha_{k,j},\alpha_{\textnormal{o},j}) = \left( \frac{N}{N+1} \right) \left(  \mathbb{E}[\left< \alpha_{.,j} \right>_N^2] + \mathbb{E}[\alpha_{\textnormal{o},j}^2] - 2\sqrt{\mathbb{E}[\left< \alpha_{.,j} \right>_N^2]\mathbb{E}[\alpha_{\textnormal{o},j}^2]} \textnormal{Corr}(\left< \alpha_{.,j} \right>_N, \alpha_{\textnormal{o},j}) \right)
\end{equation}

where we have exploited that $\mathbb{E}[\left< \alpha_{.,j} \right>_N] = \mathbb{E}[\alpha_{\textnormal{o},j}] = 0$, $\textnormal{Var}(\left< \alpha_{.,j} \right>_N) = \mathbb{E}[\left< \alpha_{.,j} \right>_N^2]$, and $\textnormal{Var}( \alpha_{\textnormal{o},j} ) = \mathbb{E}[\alpha_{\textnormal{o},j}^2]$. The unbiased estimators presented in sections 2.1 and 2.2 are then derived by finding equivalent expressions in terms of $\mathbb{E}[\left< a_{.,j} \right>_N^2]$, $\mathbb{E}[a_{\textnormal{o},j}^2]$, $\mathbb{E}[\left< b_{.,j} \right>_N^2]$, and $\mathbb{E}[b_{\textnormal{o},j}^2]$. It is now convenient to define the climatological sample mean from section 2.1 as $\overline {x} = \frac{1}{M}\sum^{M}_{j=1}\left< x_{.,j} \right>_N$. We then start from the definition of $\mathbb{E}[\left< x_{.,j} \right>_N^2]$, which can be written equivalently in terms of $\alpha_{k,j}$ or $a_{k,j}$.

\begin{equation}    
    \mathbb{E}[\left< x_{.,j} \right>_N^2] = \mathbb{E}[ ( \left< \alpha_{.,j} \right>_N + \mu )^2 ] = \mathbb{E}[ ( \left< a_{.,j} \right>_N + \overline{x} )^2 ]
\end{equation}
which can be expanded and simplified as follows

\begin{equation}
    \begin{split}  
        \mathbb{E}[ \left< \alpha_{.,j} \right>_N^2] + \mu^2 &= \mathbb{E}[ \left< a_{.,j} \right>_N^2] + \mathbb{E}[ \overline{x}^2] + 2\mathbb{E}[\left< a_{.,j} \right>_N \overline{x} ]\\
    %<REMOVED_FOR_BREVITY>
    % &= \mathbb{E}[ \left< a_{.,j} \right>_N^2] + \mathbb{E}[ \overline{x}^2] + 2 \textnormal{Cov}(\left< a_{.,j} \right>_N, \overline{x}) + 2\mathbb{E}[\left< a_{.,j} \right>_N] \mathbb{E}[\overline{x} ]\\
    %</REMOVED_FOR_BREVITY>
    &= \mathbb{E}[ \left< a_{.,j} \right>_N^2] + \mathbb{E}[ \overline{x}^2]
    \end{split}
\end{equation}
where we have noted that $\mathbb{E}[\left< a_{.,j} \right>_N \overline{x} ]=0$. This yields
\begin{equation}    
    \mathbb{E}[ \left< \alpha_{.,j} \right>_N^2] = \mathbb{E}[ \left< a_{.,j} \right>_N^2] + \textnormal{Var}(\overline{x}).
\end{equation}
If the ensemble mean forecasts contributing to the reforecast climatological mean are independent, then we can make the following substitution

\begin{equation}
    \begin{split}  
    \textnormal{Var}(\overline{x}) &= \frac{1}{M^2}\textnormal{Var}(\left< x_{.,1} \right>_N + ... + \left< x_{.,M} \right>_N) \\
    &= \frac{1}{M}\textnormal{Var}(\left< x_{.,j} \right>_N) \\
    &= \frac{1}{M}\mathbb{E}[\left< \alpha_{.,j} \right>_N^2]
\end{split}
\end{equation}
to give 
\begin{equation}    
    \mathbb{E}[ \left< \alpha_{.,j} \right>_N^2] = \frac{M}{M-1} \mathbb{E}[ \left< a_{.,j} \right>_N^2].
\end{equation}
This relationship is equally valid for observational anomalies, which can be considered equivalent to the case where  N=1, such that 

\begin{equation}    
    \mathbb{E}[ \alpha_{\textnormal{o},j}^2] = \frac{M}{M-1} \mathbb{E}[ a_{\textnormal{o},j}^2].
\end{equation}
These results can be combined to derive an expression for the mean squared error in terms of of $\mathbb{E}[\left< a_{.,j} \right>_N^2]$ and $\mathbb{E}[a_{\textnormal{o},j}^2]$ that is unbiased for small values of $N$ and $M$. 
\begin{equation}
    \textnormal{MSE}(\alpha_{k,j},\alpha_{\textnormal{o},j}) = \left( \frac{M}{M-1} \right) \left( \frac{N}{N+1} \right) \left( \mathbb{E}[\left< a_{.,j} \right>_N^2] + \mathbb{E}[a_{\textnormal{o},j}^2] - 2\sqrt{\mathbb{E}[\left< a_{.,j} \right>_N^2]\mathbb{E}[a_{\textnormal{o},j}^2]} \textnormal{Corr}(\left< a_{.,j} \right>_N, a_{\textnormal{o},j}) \right)
\end{equation}
The equivalent unbiased estimator for anomalies calculated using method B now follows trivially by making the substitutions from appendix B.

\begin{equation}
    \left<a_{.,j}\right>_{N} = \frac{M-1}{M} \left<b_{.,j}\right>_{N}   
\end{equation}
which gives

\begin{equation}    
    \mathbb{E}[ \left< \alpha_{.,j} \right>_N^2] = \frac{M-1}{M} \mathbb{E}[ \left< b_{.,j} \right>_N^2].
\end{equation}
This result can now be used to derive an expression for the unbiased mean squared error in terms of $\mathbb{E}[\left< b_{.,j} \right>_N^2]$ and $\mathbb{E}[b_{\textnormal{o},j}^2]$. 

\begin{equation}
    \textnormal{MSE}(\alpha_{k,j},\alpha_{\textnormal{o},j}) = \left( \frac{M-1}{M} \right) \left( \frac{N}{N+1} \right) \left( \mathbb{E}[\left< b_{.,j} \right>_N^2] + \mathbb{E}[b_{\textnormal{o},j}^2] - 2\sqrt{\mathbb{E}[\left< b_{.,j} \right>_N^2]\mathbb{E}[b_{\textnormal{o},j}^2]} \textnormal{Corr}(\left< b_{.,j} \right>_N, b_{\textnormal{o},j}) \right)
\end{equation}

\section{Unbiased total anomaly variance estimates}
Following a similar process to appendix C, we can also define unbiased expressions for the total anomaly variance, where $\textnormal{Var}(x_{k,j}) = \mathbb{E}[(\alpha_{k,j} + \mu )^2] - \mu^2$ and thus $\textnormal{Var}(\alpha_{k,j}) = \mathbb{E}[\alpha_{k,j} ^2]$. We start from the definition of $\mathbb{E}[x_{k,j}^2]$, which can be written equivalently in terms of $\alpha_{k,j}$ or $a_{k,j}$ to give

\begin{equation}    
    \mathbb{E}[ (\alpha_{k,j} + \mu )^2 ] = \mathbb{E}[ ( a_{k,j} + \overline{x} )^2 ],
\end{equation}
which, noting that $\mathbb{E}[a_{k,j}\overline{x}]=0$, can be expanded and simplified as follows
\begin{equation}    
    \mathbb{E}[ \alpha_{k,j} ^2] + \mu^2 = \mathbb{E}[ a_{k,j} ^2] + \mathbb{E}[ \overline{x}^2],
\end{equation}
and again to give

\begin{equation}    
    \mathbb{E}[ \alpha_{k,j} ^2] = \mathbb{E}[ a_{k,j} ^2] + \textnormal{Var}(\overline{x}).
\end{equation}
If the ensemble mean forecasts contributing to the reforecast climatological mean are independent, then we can make the same substitution we made for the ensemble mean anomalies to give
\begin{equation}
    \begin{split}  
        \mathbb{E}[ \alpha_{k,j} ^2] &= \mathbb{E}[ a_{k,j} ^2] + \frac{1}{M}\mathbb{E}[\left< \alpha_{.,j} \right>_N^2] \\
        &= \mathbb{E}[ a_{k,j} ^2] + \frac{1}{M-1}\mathbb{E}[\left< a_{.,j} \right>_N^2],
\end{split}
\end{equation}
which can be expressed equivalently as variances

\begin{equation}    
    \textnormal{Var}(\alpha_{k,j}) = \textnormal{Var}(a_{k,j}) + \frac{1}{M-1} \textnormal{Var}(\left< a_{.,j} \right>_N).
\end{equation}
Noting that ensemble spread is equivalent using methods A and B such that $\textnormal{Var}(a_{k,j}) - \textnormal{Var}(\left< a_{.,j} \right> ) =  \textnormal{Var}(b_{k,j}) - \textnormal{Var}(\left< b_{.,j} \right> )$, we arrive at equivalent expressions for the total variance using method B

\begin{align}    
    \mathbb{E}[ \alpha_{k,j} ^2] &= \mathbb{E}[ b_{k,j} ^2] - \frac{1}{M}\mathbb{E}[\left< b_{.,j} \right>_N^2],\\
    \textnormal{Var}(\alpha_{k,j}) &= \textnormal{Var}(b_{k,j}) - \frac{1}{M} \textnormal{Var}(\left< b_{.,j} \right>_N).
\end{align}
In the case of a single member (i.e. N=1), these expressions are equivalent to those for ensemble mean anomalies derived above.

\section{Unbiased member-by-member calibration}
\citet{johnson2009reliability} examine the statistical properties of a simple member-by-member calibration approach that is appropriate for ensemble forecast anomalies (or unbiased forecasts) and widely used in seasonal forecasting. They demonstrate that the resulting calibrated forecasts satisfy both the climatological and ensemble variance reliability criteria described in section \ref{section:methods}.  However, they do not account for the effects of a finite ensemble size. The member-by-member approach separately modifies the ensemble mean and perturbations from the ensemble mean as follows

\begin{equation}
\hat{z}_{k,j} = \kappa \left<z_{.,j}\right>_N + \lambda  \left[ z_{k,j} - \left<z_{.,j}\right>_N   \right],
\end{equation}

where $z_{k,j}$ represents an uncalibrated ensemble member anomaly, $\left<z_{.,j}\right>_N$ represents an uncalibrated ensemble mean anomaly, $\hat{z}_{k,j}$ represents a calibrated ensemble member anomaly, and $\kappa$ and $\lambda$ are parameters to be estimated. To simplify the following algebra and comparisons with \citet{johnson2009reliability}, we make the following substitutions:

\begin{align}
    \sigma_{\textnormal{o}}^2 = \mathbb{E}[z_{\textnormal{o},j}^2],\\
    \sigma_{z}^2 = \mathbb{E}[z_{k,j}^2],\\
    \sigma_{\left<z\right>}^2 = \mathbb{E}[\left<z_{.,j}\right>_N ^2],\\
    \sigma_{s}^2 = \mathbb{E}[(z_{k,j} - \left<z_{.,j}\right>_N)^2 ],\\
    \label{eq:calibrated_variance} \sigma_{\hat{z}}^2 = \kappa^2 \sigma_{\left<z\right>}^2 + \lambda^2 \sigma_{s}^2,\\ 
    \rho = \textnormal{Corr}(\left<z_{.,j}\right>_N, z_{\textnormal{o},j}) = \textnormal{Corr}(\left<\hat{z}_{.,j}\right>_N, z_{\textnormal{o},j}),
\end{align}
where expectations are taken over ensemble members and start dates, $\sigma_{\textnormal{o}}^2$ is the climatological variance of observed anomalies, $\sigma_{z}^2$ is the climatological variance of uncalibrated ensemble member anomalies, $\sigma_{\left<z\right>}^2$ is the climatological variance of uncalibrated ensemble mean anomalies, $\sigma_{s}^2$ is the average ensemble variance of uncalibrated anomalies, $\sigma_{\hat{z}}^2$ is the climatological variance of calibrated ensemble member anomalies, and $\rho$ is the correlation between the adjusted (or unadjusted) ensemble mean and the observed truth. 

\citet{johnson2009reliability} estimate the parameters $\kappa$ and $\lambda$ by enforcing the following constraints: (i) the variance of calibrated members should be equal to the variance of the truth (i.e. $\sigma_{\textnormal{o}}^2 = \sigma_{\hat{z}}^2$) and (ii) the correlation of the calibrated members with the ensemble mean should be the same as the correlation of the truth with the ensemble mean (i.e. $\textnormal{Corr}(\left<\hat{z}_{.,j}\right>_N, \hat{z}_{k,j}) = \rho$). This provides the following estimates for $\kappa$ and $\lambda$

\begin{align}
    \label{eq:kappa} \kappa = \rho \frac{\sigma_{\textnormal{o}}}{\sigma_{\left<z\right>}}, \\   
    \lambda^2  = (1 - \rho^2)\frac{\sigma_{\textnormal{o}}^2}{\sigma_{s}^2}
\end{align}

These choices of $\kappa$ and $\lambda$ implicitly satisfies two further conditions: (i) the RMSE of ensemble mean anomalies is minimized and (ii) the average ensemble variance converges with the average squared error of the ensemble mean when averaged over many cases (i.e.  $\frac{\textnormal{Spread}}{\textnormal{RMSE}}  \rightarrow 1 \textnormal{ as }M \rightarrow \infty$), regardless of ensemble size. This latter condition is inconsistent with our expectations for a perfectly reliable ensemble forecast of finite ensemble size (see equation \ref{eq:spread_error}). The main consequence of this bias is that forecasts calibrated using this method are overdispersive for small ensemble sizes. 

Here, we present a generalisation that results in calibrated forecasts with unbiased spread-error ratios, even for small ensemble sizes. This is achieved by constraining calibrated forecasts using the same climatological reliability criterion but enforcing an unbiased spread-error ratio for finite ensemble sizes instead of a correlation-based constraint. We start from equation \ref{eq:spread_error} expressed in terms of the mean squared error and average ensemble variance of the calibrated anomalies: 

\begin{equation}
    \left( \frac{\textnormal{Spread}}{\textnormal{RMSE}} \right)^2 = \frac{\lambda^2 \sigma_{s}^2R}{\mathbb{E}[ ( \kappa \left<z_{.,j}\right>_N - z_{\textnormal{o},j} )^2 ]} = 1
\end{equation}      

where $R = \frac{N+1}{N-1}$ ensures estimates are unbiased with ensemble size ($N$). Rearranging and expanding gives

\begin{equation}
\lambda^2 \sigma_{s}^2R =  \kappa^2 \sigma_{\left<z\right>}^2 + \sigma_{\textnormal{o}}^2 - 2 \kappa \rho \sigma_{\textnormal{o}}\sigma_{\left<z\right>}.
\end{equation}
We then eliminate ensemble variance using equation \ref{eq:calibrated_variance} and enforce the constraint that the variance of calibrated members should be equal to the variance of the truth (i.e. $\sigma_{\textnormal{o}}^2 = \sigma_{\hat{z}}^2$) to get the following quadratic expression in $\kappa$  

\begin{equation}
    \kappa^2 \sigma_{\left<z\right>}^2 (R + 1) - 2 \kappa \rho \sigma_{\textnormal{o}}\sigma_{\left<z\right>} + \sigma_{\textnormal{o}}^2(1-R).
\end{equation}
This equation has two real roots, one of which is the trivial solution given by $\kappa \rightarrow 0$ as N $\rightarrow \infty$. The non-trivial solution is given by 

\begin{equation}
    \label{eq:unbiased_kappa}
    %<REMOVED_FOR_BREVITY>
    %    \kappa =  \frac{2 \rho \sigma_{\textnormal{o}}\sigma_{\left<z\right>} + \sqrt{4 \rho^2  \sigma_{\textnormal{o}}^2 \sigma_{\left<z\right>}^2 + 4 \sigma_{\textnormal{o}}^2 \sigma_{\left<z\right>}^2 (R+1)(R-1)}}{2\sigma_{\left<z\right>}^2(R+1)},
    %</REMOVED_FOR_BREVITY>
   \kappa = \frac{\sigma_{\textnormal{o}}}{\sigma_{\left<z\right>}} \left( \frac{\rho + \sqrt{\rho^2 + R^2 - 1}}{R+1} \right)
\end{equation}

with $\lambda$ calculated by substituting $\kappa$ into equation \ref{eq:calibrated_variance}. From inspection, it is clear that equations \ref{eq:unbiased_kappa} and \ref{eq:kappa} converge as $N \rightarrow \infty$ and $R \rightarrow 1$. 
\end{appendices}
   
%==========================
%   R E F E R E N C E S 
%==========================
\newpage
\bibliographystyle{rss}
\bibliography{references}

%==========================
% F I G U R E S 
%==========================
\newpage

%============================
% S U P P .  F I G U R E S 
%============================
\renewcommand{\thefigure}{S1}
\begin{figure}[!htbp]
\includegraphics[width=15cm]{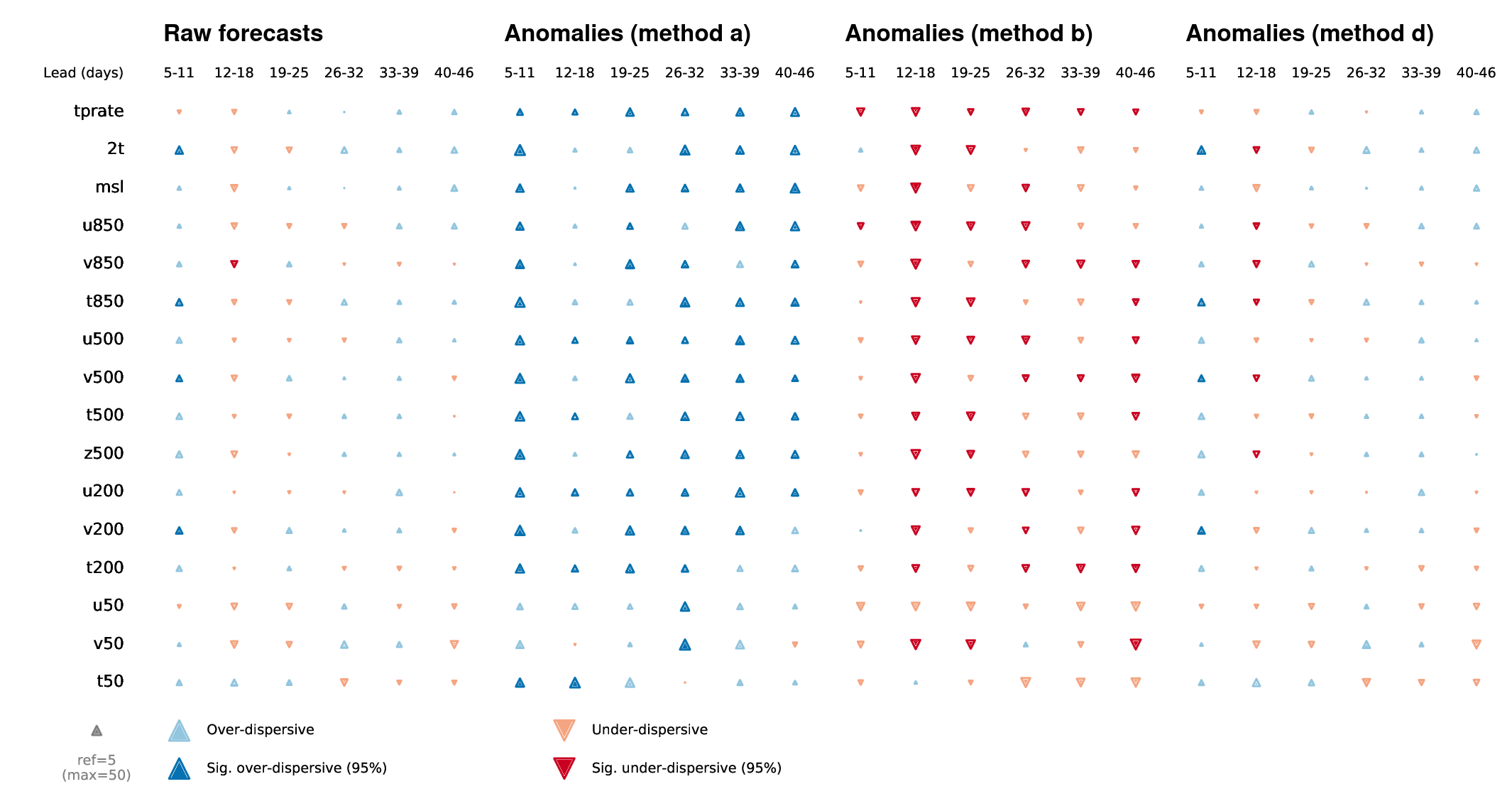}
\caption{As figure 3 in the main text, but for a perfectly reliable ensemble constructed from perturbed members 1 to 5 for the reforecast period 2001-2020 (i.e. $N$=5, $M$=20).}
\end{figure}

\renewcommand{\thefigure}{S2}
\begin{figure}[!htbp]
    \includegraphics[width=8cm]{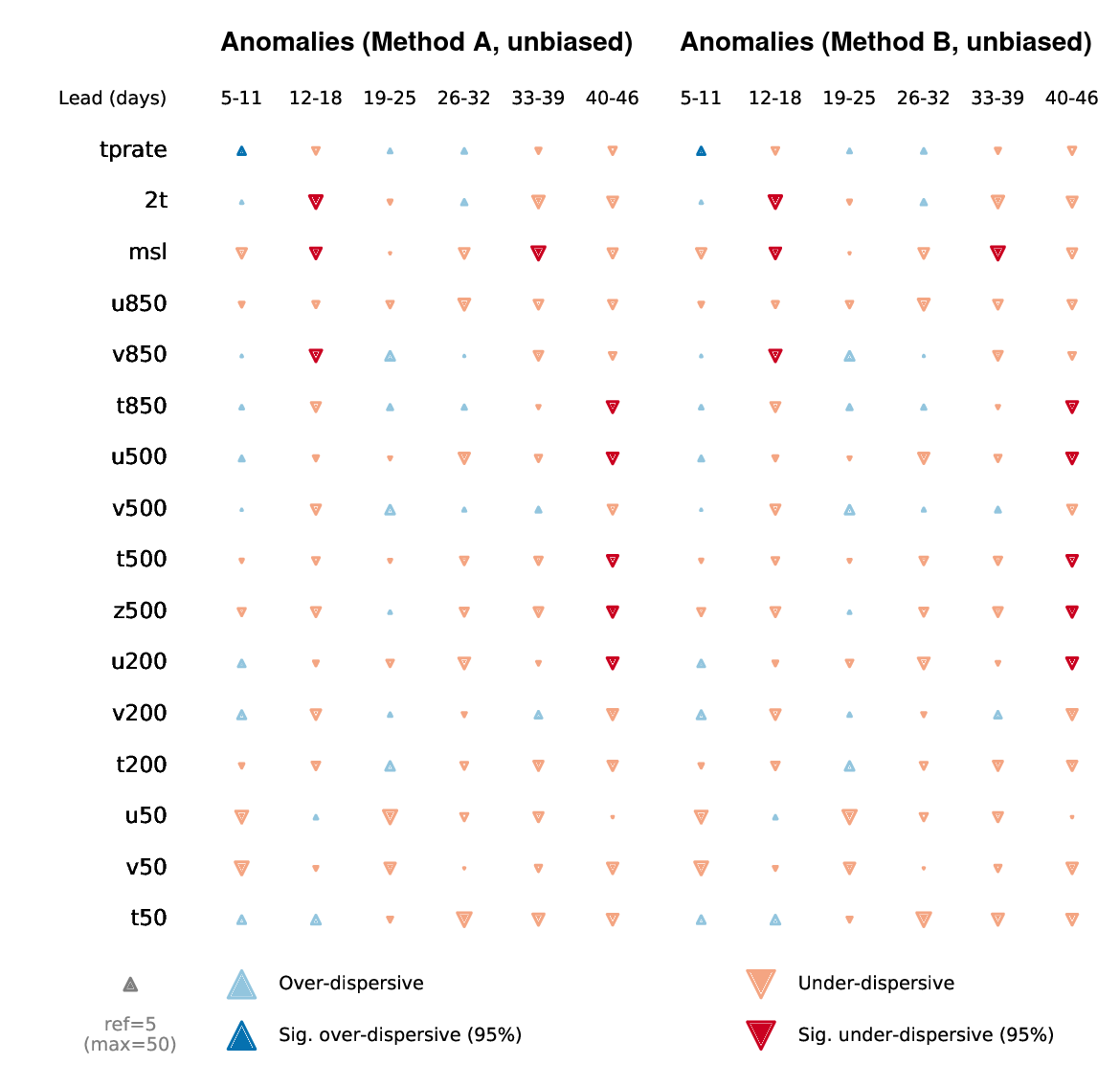}
    \centering
    \caption{As figure 5 in the main text, but for the reforecast period 2016-2020 (i.e. $N$=9, $M$=5).}
\end{figure}

\end{document}